\documentclass[lettersize,journal]{IEEEtran}
\usepackage{amsmath,amsfonts}
\usepackage{algorithmic}
\usepackage{algorithm}
\usepackage{array}
\usepackage[caption=false,font=normalsize,labelfont=sf,textfont=sf]{subfig}
\usepackage{textcomp}
\usepackage{stfloats}
\usepackage{url}
\usepackage{verbatim}
\usepackage{graphicx}
\usepackage{cite}
\usepackage{bm}
\usepackage{relsize}
\DeclareMathOperator*{\A}{ \mathlarger{\mathlarger{\mathlarger{\boldsymbol{\mathsf{A}}}}} }

\begin{document}

\title{A Practical Finite Element Approach for Simulating Dynamic Crack Growth in Cu/Ultra Low-k Interconnect Structures}

\author{Yuxi Xie$^{*}$, Ethan J. Wu, Lu Xu, Jimmy Perez, Shaofan Li
\thanks{This paper was produced by the IEEE Publication Technology Group. They are in Piscataway, NJ.}
\thanks{Manuscript received XX XX, XXXX; revised XX xx, XXXX.}}

\markboth{Journal of \LaTeX\ Class Files,~Vol.~14, No.~8, August~2021}%
{Shell \MakeLowercase{\textit{et al.}}: A Sample Article Using IEEEtran.cls for IEEE Journals}


\maketitle

\begin{abstract}
This work presents a practical finite element modeling strategy, the Crack Element Method (CEM), for simulating the dynamic crack propagation in two-dimensional structures. The method employs an element-splitting algorithm based on the Edge-based Smoothed Finite Element Method (ES-FEM) to capture the element-wise crack growth while reducing the formation of poorly shaped elements that can compromise numerical accuracy and computational performance. A fracture energy release rate formulation is also developed based on the evolving topology of the split elements. The proposed approach is validated through a series of classical benchmark problems, demonstrating its accuracy and robustness in addressing dynamic fracture scenarios. Finally, the applicability of the CEM is illustrated in a case study involving patterned Cu/Ultra Low-k interconnect structures.
\end{abstract}


\section{Introduction}
\IEEEPARstart{T}{he} initiation and propagation of cracks under transient dynamic loading are widespread phenomena that can cause significant issues in interconnects within integrated circuits (ICs), leading to electrical failures, reliability problems, and reduced performance. Compared to quasi-static loading, including inertia effects significantly complicates the mechanisms of crack initiation and propagation\cite{ravi1984experimental}. This complexity arises from several contributing factors, including the strong stress field nonlinearity, particularly the singularity near the crack tip, material heterogeneity, variations in stress wave propagation within the structure, and wave reflections at boundaries. Additionally, the degree of plasticity involved in crack development reflects the material’s brittleness or ductility, further complicating the problem\cite{anderson2005fracture}. As a result, dynamic fracture in solids has emerged as one of the most challenging and critical topics in solid mechanics over the past few decades. It not only presents fundamental scientific uncertainties but also holds substantial commercial relevance, particularly in fields such as microelectronics.

The scaling law in the semiconductor industry continues to impose increasingly strict demands on chip design, manufacturing, and packaging technologies. Size reduction and performance enhancement drive high-volume manufacturing to satisfy both aspects of Moore’s Law. One example is adopting the Cu/low-k integration scheme in the Backend-of-Line (BEoL) layers of advanced CMOS technology nodes. However, as device dimensions shrink, the thermomechanical and chemical reliability of low-k dielectric thin films deteriorates, posing a significant challenge. The mechanical integrity of BEoL structures is compromised by low-k and ultralow-k (ULK) materials due to their inherently porous architecture and low fracture toughness\cite{grill2014progress}. In addition to mechanical failures, mismatches in the coefficients of thermal expansion (CTE) between various packaging materials and the silicon die induce substantial thermomechanical stress. This chip-package interaction (CPI) further undermines long-term device reliability. Failures in BEoL stacks generally fall into two categories: (1) delamination along the Cu/dielectric interface and (2) fracture within the low-k dielectric materials. Delamination, an adhesive failure, often results from weak Cu/dielectric adhesion caused by thermal mismatch, high current-induced atomic migration, moisture ingress, and residual stress from processes such as chemical mechanical polishing (CMP) and packaging. In contrast, fracture within the dielectric, classified as a cohesive failure, typically initiates due to the porous nature of low-k materials and high residual stresses introduced during wafer bonding, CMP, and packaging steps.

CPI has become an area of increasing concern, particularly when stiff interconnect components, such as lead-free solder balls and copper (Cu) pillars, are integrated with thin-film materials like porous organosilicate glass (OSG), which exhibit low-k dielectric properties\cite{wang2005chip}. Various techniques have been developed to mitigate or even eliminate the adverse effects of CPI. Among them, integrating metallic guard ring (GR) structures into the BEoL stack has emerged as one of the most promising approaches. The primary function of the GR structure is to absorb and dissipate externally induced energy, thereby reducing or preventing the initiation and propagation of cracks\cite{zhang2009chip}. A comprehensive study of the kinematics involved in the dynamic fracture process, particularly in the presence of mitigation measures such as GR structures, is crucial for identifying vulnerable regions within the BEoL stack. This understanding is essential for enhancing the mechanical robustness and long-term reliability of on-chip interconnects.

To fully elucidate the mechanisms of dynamic crack initiation and propagation, extensive efforts have been dedicated to numerous in-situ experiments, analytical models, and numerical methods based on a variety of theoretical frameworks\cite{song2008comparative,wu2016meshfree,kalthoff2000modes,cox2005modern,doan2017rate}. Compared to costly experiments and the inherent limitations of analytical models\cite{griffith1921vi,schapery1975theory,chudnovsky1999application}, numerical methods have emerged as the primary tools for investigating dynamic fracture problems. Their capacity to model arbitrary solid geometries with complex boundary conditions offers significant advantages in terms of flexibility, computational efficiency, and simulation accuracy. Numerical approaches to fracture modeling are typically categorized into two main classes: discrete crack approaches and smeared crack approaches. The discrete crack approach introduces displacement discontinuities along explicitly modeled cracks once failure criteria are met. This method is commonly implemented within the finite element framework\cite{ngo1967finite,nilson1968nonlinear}. In this approach, cracks are simulated through element splitting, with new nodes and element edges introduced to represent crack surfaces. As a result, crack propagation is constrained by mesh orientation, making the accuracy of this method highly dependent on the mesh quality. In contrast, the smeared crack approach distributes damage or cracking over a region of the finite element mesh, smoothing material degradation across several elements. Since its initial application in modeling cracks in prestressed concrete vessels\cite{rashid1968ultimate}, this approach has gained popularity, particularly for quasi-brittle materials, due to its compatibility with standard finite element programs and its lack of restriction on crack directionality. However, the smeared crack method suffers from significant mesh dependency in energy dissipation and crack path prediction, which continues to limit its practical applicability\cite{cedolin1980effect,pietruszczak1980numerical}. 

This work presents a Crack Element Method (CEM) for simulating transient dynamic crack propagation in quasi-brittle materials. Built upon the Edge-based Smoothed Finite Element Method (ES-FEM) framework, the proposed element-splitting algorithm effectively captures the evolution of stress fields under dynamic loading with accuracy and efficiency. Unlike conventional finite element approaches, where cracks are introduced via additional nodes and split element edges (discrete methods) or through smeared damage across neighboring elements (smeared methods), CEM avoids constraints related to mesh orientation and coarseness. This flexibility promotes computational simplicity, enables straightforward implementation, and supports smooth extension to three-dimensional problems. In parallel, a novel and intuitive failure criterion is introduced, derived from the topology of split elements and a local computation of the fracture energy release rate. In contrast to traditional approaches, such as the J-integral, often limited by mesh sensitivity and the challenges of crack tip tracking under transient dynamic conditions, the proposed criterion offers greater robustness and objectivity. By adopting an energy-based formulation, the CEM effectively avoids common issues like crack path trapping caused by stress wave reflections and oscillations. As a locally tracking crack algorithm, CEM offers an effective balance between accuracy, robustness, and computational practicality.

The paper is structured as follows to provide a comprehensive understanding of the proposed methodology: Section $\mathrm{II}$ reviews the fundamentals of ES-FEM and the variational formulation of two-dimensional fracture problems in solid mechanics applications. Section $\mathrm{III}$ presents the proposed Crack Element Method, including a novel approach for computing the fracture energy release rate. Section $\mathrm{IV}$ validates the method through several classic benchmark problems involving transient dynamic crack propagation in quasi-brittle materials and advanced cases of mechanically and thermally induced crack propagation relevant to Cu/Ultra Low-k interconnect systems. Finally, Section $\mathrm{V}$ concludes the study, discusses current limitations, and outlines future directions for further development and application of the method.

\section{Problem Formulation of 2D Fractures based on ES-FEM}
This section offers a brief overview of the two-dimensional fracture problem along with its mathematical formulation. It also outlines the fundamental principles of the Edge-based Smoothed Finite Element Method for transient dynamic simulations. Some basic notations and indices are also introduced. 

\subsection{Variational principle on 2D fractures}
\begin{figure}[!t]
\centering
\includegraphics[width=2.5in]{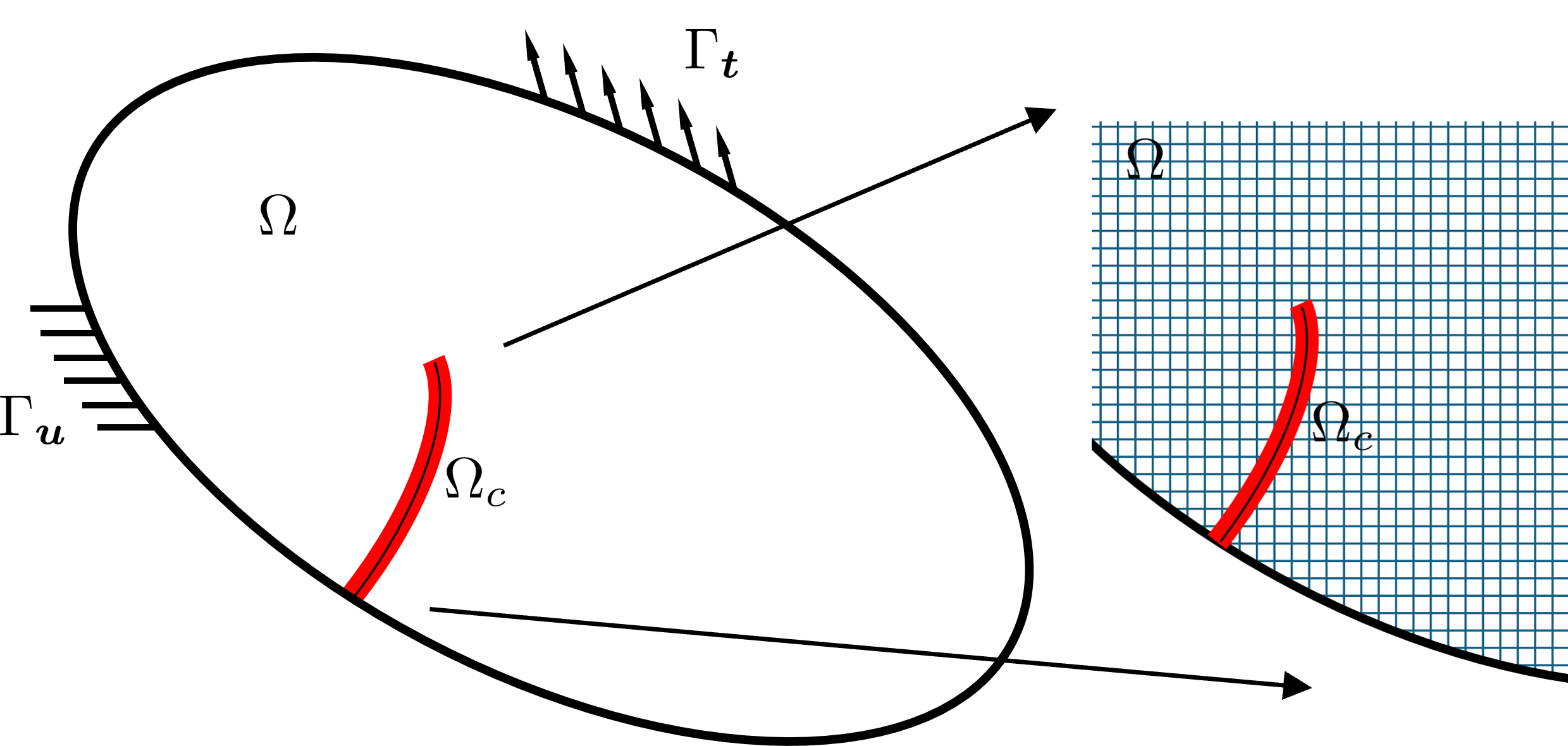}
\caption{Illustration of concepts and notations in Lagrangian dynamics and Zoom-in of the cracked domain with grid discretization.}
   \label{fig21:Lagrangian_illustration}
\end{figure}

Based on the Lagrangian description for a dynamic system as follows,
\begin{equation}
\mathcal{L} = \mathcal{T} - \mathcal{W}_{int} + \mathcal{W}_{ext}
\end{equation}
in which, $\mathcal{T}$ indicates the kinetic energy of the dynamic system, $\mathcal{W}_{int}$ indicates strain gradient energy density, and $\mathcal{W}_{ext}$ indicates external potentials.   \\
In a two-dimensional fracture problem, the three terms above can be characterized as following equations,
\begin{eqnarray}
\mathcal{T} &=& \frac{1}{2} \int_{\Omega} \rho \dot{\bm{u}} \cdot \dot{\bm{u}} d \Omega \label{eq:Lagrangian_kinetic} \\
\mathcal{W}_{int} &=& \frac{1}{2} \int_{\Omega} \bm{\varepsilon}: \bm{\sigma} d \Omega \label{eq:Lagrangian_internal} \\
\mathcal{W}_{ext} &=& \int_{\Omega} \bm{u} \cdot \bm{b} d\Omega + \int_{\Gamma_t} \bm{u} \cdot \bar{\bm{t}} d\Gamma \label{eq:Lanrangian_external}
\end{eqnarray}
in which, $\dot{\bm{u}}\left(\bm{x}\right)$ represents the velocity of the studied material point $\bm{x}$ and $:$ indicates two indices contraction. $\Omega$ represents the studied domain while $\Omega_c$ represents the cracked domain marked by a red line in Figure.\ref{fig21:Lagrangian_illustration}. The Dirichlet and Neuman boundary conditions are represented by $\Gamma_u$ and $\Gamma_t$, respectively and $\rho$ is the material density of the domain $\Omega$. It is noteworthy that $\bm{\sigma} = \bm{\sigma}\left(\bm{u}, \bm{x} \right)$, i.e.,
\begin{equation}
\bm{\sigma} = \begin{cases}
\bm{\sigma}\left(\bm{x} \right), & \bm{x} \in \Omega \backslash \Omega_c \\
\bm{0}, & \bm{x} \in \Omega_c
\end{cases}
\end{equation}

The strain energy stored in the Lagrangian system, as stated in Eq.(\ref{eq:Lagrangian_internal}), arises from the work done by internal stresses acting on the strain field within the discussed domain. In this context, $\bm{\varepsilon}$ denotes the infinitesimal strain tensor, while, the associated stress tensor, $\bm{\sigma}$ is in terms of $\bm{\varepsilon}$. For example, in the linear elastic case $\bm{\sigma} = \mathbb{C} : \bm{\varepsilon}$, $\mathbb{C}$ refers to the constitutive tensor. The strain tensor is specifically defined as the symmetric part of the gradient of the displacement field, i.e., $\bm{\varepsilon} = \nabla^{s} \bm{u}$. Additionally, Eq.(\ref{eq:Lanrangian_external}) defines the potential associated with external loads, where $\bm{b}$ and $\bar{\bm{t}}$ correspond to body forces and surface tractions, respectively.

Based on Hamilton's principle, the actual trajectory of a system makes the time integral of the Lagrangian stationary, i.e., $\int_{t_1}^{t_2} \mathcal{L} dt = 0$. The displacement field is perturbed by an admissible variation $\delta \bm{u}$ that satisfies the essential/Dirichlet boundary conditions. Furthermore, it is assumed that the varied trajectory may deviate from the actual one, except at the initial and final time stages, $t_1$ and $t_2$. As a mathematical description:
\begin{eqnarray}
&&\delta \bm{u} = \bm{0}, \text{ if } \bm{x} \in \Gamma_{\bm{u}} \\
&&\delta \bm{u}\left(\bm{x}, t_1 \right) = \delta \bm{u} \left( \bm{x}, t_2 \right) = \bm{0}, \text{ if } \bm{x} \in \Omega
\end{eqnarray}

Therefore, the variation of the Lagrangian time integration is written as follows,
\begin{eqnarray}
\delta \mathcal{L} &=& \delta \left( \mathcal{T} - \mathcal{W}_{int} + \mathcal{W}_{ext} \right) \Rightarrow \nonumber \\
 \delta \int_{t_1}^{t_2} \mathcal{L} dt &=& \int_{t_1}^{t_2} \Big( \rho \dot{\bm{u}} \cdot \delta \dot{\bm{u}} d \Omega - \int_{\Omega \backslash \Omega_c}  \nabla^{s} \delta \bm{u} : \bm{\sigma} d \Omega \nonumber \\
&&+ \int_{\Omega} \delta \bm{u} \cdot \bm{b} d\Omega + \int_{\Gamma_{\bm{t}}} \delta \bm{u} \cdot \bar{\bm{t}} d\Gamma \Big) dt = 0 \label{eq:Lagrangian_variation}
\end{eqnarray}
By exchanging the integral order of time and space on kinetic energy variation, it is demonstrated as follows,
\begin{eqnarray}
&&\int_{t_1}^{t_2} \int_{\Omega} \rho \dot{\bm{u}} \cdot \delta \dot{\bm{u}} d\Omega dt = \int_{\Omega} \left(\int_{t_1}^{t_2} \rho \dot{\bm{u}} \cdot \delta \dot{\bm{u}} dt \right) d\Omega \nonumber \\
&&= \int_{\Omega} \left( \rho \dot{\bm{u}} \cdot \delta \bm{u}\Big|_{t_1}^{t_2} - \int_{t_1}^{t_2} \rho \ddot{\bm{u}} \cdot \delta \bm{u} dt \right) d\Omega \nonumber \\
&&= -\int_{t_1}^{t_2} \int_{\Omega} \rho \ddot{\bm{u}} \cdot \delta \bm{u} d\Omega dt \label{eq:kinetic_intorder_change}
\end{eqnarray}
since $\rho \dot{\bm{u}} \cdot \delta \bm{u} \Big|_{t_1}^{t_2} = \rho \dot{\bm{u}} \cdot \left(\delta\bm{u}_2 - \delta \bm{u}_1 \right) = 0$. Substituting Eq.(\ref{eq:kinetic_intorder_change}) into Eq.(\ref{eq:Lagrangian_variation}), the variation of Lagrangian time integration can be converted into a temporal-independent equation as follows,
\begin{eqnarray}
\int_{\Omega} \rho \ddot{\bm{u}} \cdot \delta \bm{u} d\Omega &=&-\int_{\Omega \backslash \Omega_c} \nabla^s \delta \bm{u} : \bm{\sigma} d\Omega + \int_{\Omega} \delta \bm{u} \cdot \bm{b} d\Omega \nonumber \\
&&+ \int_{\Gamma_{\bm{t}}} \delta \bm{u}\cdot \bar{\bm{t}} d\Gamma \label{eq:weak_form}
\end{eqnarray}

\subsection{Edge-based Smoothed FEM discretization}
\begin{figure*}[!t]
\centering
\subfloat[]{\includegraphics[width=2.0in]{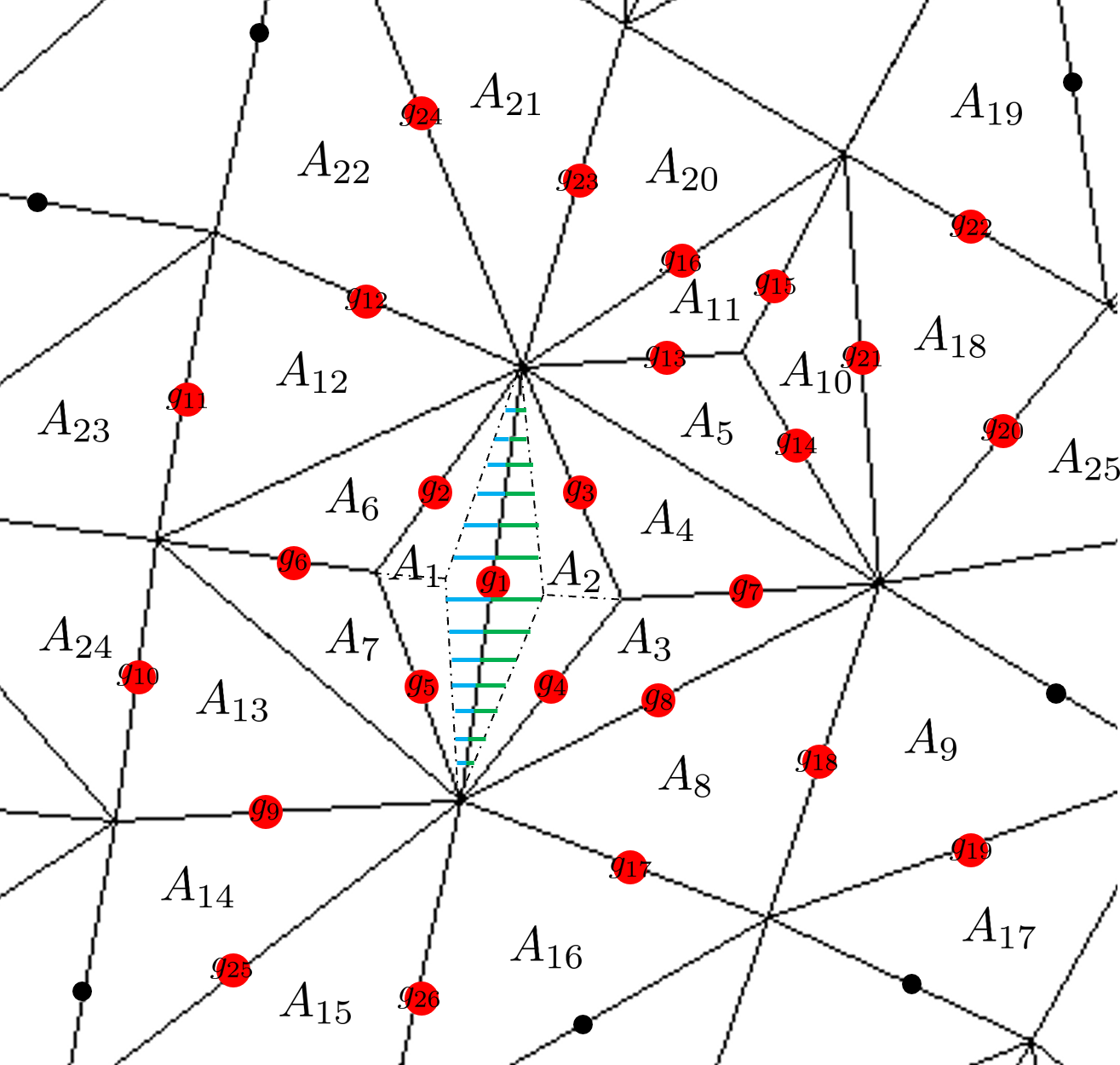}}
\hfil
\subfloat[]{\includegraphics[width=2.0in]{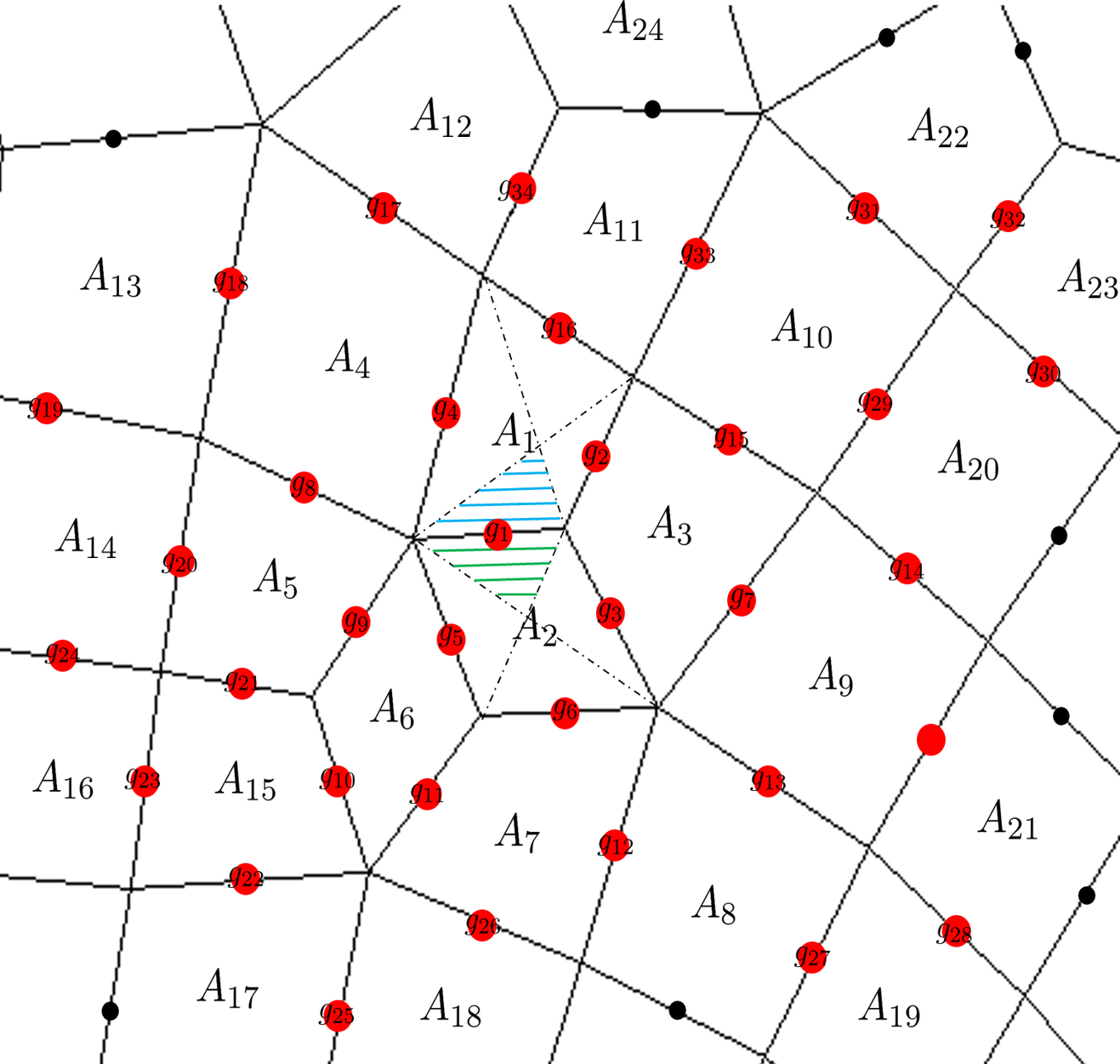}}
   \caption{The studied domain is discretized by red points while black points are skipped. Notation $A_i$ represents element ID and $g_i$ represents edge quadrature ID: (a). ES-FEM formulation of Constant Strain Triangle element, totally $24$ related elements and $26$ edge quadrature points; (b). ES-FEM formulation of bilinear Quadrilateral element, totally $24$ related elements and $34$ edge quadrature points. The blue shadow and green shadow represent shared region of element $A_1$ and $A_2$ by edge quadrature point $g_1$, respectively.}
   \label{fig22:ES-FEM_CST_QUAD_formulation}
\end{figure*}

Edge-based Smoothed Finite Element Method (ES-FEM) was first proposed by Liu et al.\cite{liu2009edge}, in which the strain field is averaged over localized smoothing domains, shifting the computation of the stiffness matrix from a traditional element-based approach to one that relies on these smoothing domains. The elements' edges take up of whole studied domain as $\Omega = \bigcap_{i=1}^{N_{edges}}$ and $N_{edges}$ is the total number of edges of elements in the entire domain. For the commonly used Constant Strain Triangle (CST) element and bilinear Quadrilateral (QUAD) element, any one of the edges is shared by two neighboring elements, as shown in Figure.\ref{fig22:ES-FEM_CST_QUAD_formulation}.

The displacement field at the edge quadrature point $g_i$ can be interpolated as follows,
\begin{eqnarray}
\bm{u} \left(g_i \right) = \sum_{j \in \mathcal{N}^{g_i}_e } \left( \sum_{k \in \mathcal{M}_n^{A_j} } N_k^{A_j}(g_i) \bm{u}_k \times \frac{A_j}{\sum_{m \in \mathcal{L}_e^{g_i}} A_m} \right)
\label{eq:disp_ES_FEM}
\end{eqnarray}
in which, $g_i$ indicates edge quadrature point ID, $A_j$ indicates neighboring element ID, both $\mathcal{N}_e^{g_i}$ and $\mathcal{L}_e^{g_i}$ indicate the neighboring elements set of edge quadrature point $g_i$, $\mathcal{M}_n^{A_j}$ indicates the nodes set of the neighboring element $A_j$, $N_k^{A_j}\left(g_i \right)$ indicates the $k$-th nodal shape function of element $A_j$ at edge quadrature $g_i$ location, and $\bm{u}_k$ indicates the displacement of node $k$.

The infinitesimal strain field can be obtained by taking the derivative of displacement $\bm{u}\left(g_i \right)$ in terms of current coordinates $\bm{x}$, as shown below
\begin{eqnarray}
\bm{\varepsilon} &=& \frac{\partial \bm{u}\left(g_i \right)}{\partial \bm{x}} \nonumber \\
&=& \sum_{j \in \mathcal{N}^{g_i}_e } \left( \sum_{k \in \mathcal{M}_n^{A_j} } \frac{\partial N_k^{A_j}(g_i)}{\partial \bm{x}} \bm{u}_k \cdot \frac{A_j}{\sum_{m \in \mathcal{L}_e^{g_i}} A_m} \right)
\end{eqnarray}
so that the discretized matrix formulation can be obtained,
\begin{eqnarray}
\bm{\varepsilon} = \left(\varepsilon_{11}, \varepsilon_{22}, \gamma_{12} \right)^T = \A_j \left( \A_k \left[B_k^{A_j} \right]  w_{A_j}  \right) \cdot \left\{\bm{u} \right\}
\label{eq:strain_discretization}
\end{eqnarray}
in which, $j$ represents the neighboring element ID, $k$ represents the nodal ID of a neighboring element, $w_{A_j} = \frac{A_j}{\sum_n A_n}$ represents the area weight of each neighboring element, $\left[B_k^{A_j}\right]$ represents the stress-strain matrix of Node $k$ in element $A_j$ and $\A$ is the element assembly operator in FEM.\\
In the element formulation of Constant Strain Triangle or bilinear Quadrilateral, Eq.(\ref{eq:strain_discretization}) is expanded as,
\begin{eqnarray}
&&\bm{\varepsilon} = \begin{Bmatrix}
\varepsilon_{11} \\
\varepsilon_{22} \\
\gamma_{12}
\end{Bmatrix} = \nonumber \\
&&\left[ \begin{array}{cc|cc}
\frac{\partial N_i^1}{\partial x} w_1 & 0 & \frac{\partial N_i^2}{\partial x}  w_2 & 0 \\
0 & \frac{\partial N_i^1}{\partial y} w_1 & 0 & \frac{\partial N_i^2}{\partial y} w_2 \\
\frac{\partial N_i^1}{\partial y} w_1 & \frac{\partial N_i^1}{\partial x} w_1 & \frac{\partial N_i^2}{\partial y} w_2 & \frac{\partial N_i^2}{\partial x} w_2
\end{array} \right] \begin{Bmatrix}
u_i^1 \\
v_i^1 \\
u_i^2 \\
v_i^2 \\
\end{Bmatrix} ~.
\label{eq:CST_QUAD_strain_discretization}
\end{eqnarray}
in which, $N_i^j$ represents the $i$-th nodal shape function of the neighboring element $e_j$, $w_j = \frac{A_j}{\sum_k A_k }$ represents the area weight of each neighboring element, $u_i^{e_j}$ and $v_i^j$ represent $i$-th nodal displacement of the neighboring element $e_j$ along $x$ and $y$-direction, respectively. The subscript $i = 1\sim 3$ so that the dimensions of the $\bm{B}$ matrix are $3\times 12$ for the triangle element while the subscript $i = 1\sim 4$ so that the dimensions of the $\bm{B}$ matrix are $3\times 16$ for the quadrilateral element. 

It is noteworthy that the internal force $f_{int}$ assembly process in ES-FEM is slightly different from standard FEM. After the displacement of studied domain is discretized, the internal force in the Eq.(\ref{eq:weak_form}) becomes as follows,
\begin{eqnarray}
f_{int} = \int_{\Omega \backslash \Omega_c} \nabla^s \bm{N} : \bm{\sigma} d\Omega = \A_{j} \left( \A_k \left[ B_k^{A_j} \right]^T w_{A_j} \cdot \bm{\sigma} \right) \cdot A_j
\label{eq:fint_discretization}
\end{eqnarray}
in which, $j$ represents the neighboring element ID, $k$ represents nodal ID of a neighboring element, $w_{A_j} = \frac{A_j}{\sum_n A_n}$ represents the area weight of each neighboring element and $A_j$ represents the area of a neighboring element. \\
In element formulations, Eq.(\ref{eq:fint_discretization}) can be expanded as,
\begin{equation}
\begin{aligned}
&\bm{f}_{int} = \begin{Bmatrix}
f_{xi}^{e1} \\
f_{yi}^{e1} \\
f_{xi}^{e2} \\
f_{yi}^{e2}
\end{Bmatrix} =  \frac{\sum A_j}{r} \cdot \\
&\left[ \begin{array}{cc|cc}
\frac{\partial N_i^1}{\partial x} w_1, & 0, & \frac{\partial N_i^2}{\partial x}  w_2, & 0 \\
0, & \frac{\partial N_i^1}{\partial y} w_1, & 0, & \frac{\partial N_i^2}{\partial y} w_2 \\
\frac{\partial N_i^1}{\partial y} w_1, & \frac{\partial N_i^1}{\partial x} w_1, & \frac{\partial N_i^2}{\partial y} w_2, & \frac{\partial N_i^2}{\partial x} w_2
\end{array} \right]^T \begin{Bmatrix}
\sigma_{11} \\
\sigma_{22} \\
\sigma_{12}
\end{Bmatrix}
\label{eq:CST_QUAD_fint_discretization}
\end{aligned}
\end{equation}
where $f_{xi}^j$ and $f_{yi}^j$ represent the $i$-th nodal force of neighboring element $e_j$ in $x$- and $y$- direction, respectively. $\left[\cdot\right]^T$ represents the transpose of the matrix. $r=3$ for triangle element while $r=4$ for quadrilateral element.  Other notations follow definitions above as well.

Other terms in Eq.(\ref{eq:weak_form}) follow standard finite element discretization, such as inertial term $\int_{\Omega} \rho \ddot{\bm{u}} \cdot \delta \bm{u}d\Omega $ can be formulated into a diagonal mass matrix:
\begin{eqnarray}
\bm{M} &=&  \begin{bmatrix}
m_{1}, & 0, & \cdots, & 0 \\
0, & m_{2},  & \cdots, & 0 \\
\vdots & \vdots & \ddots & \vdots \\
0 & 0 & \cdots & m_{N}
\end{bmatrix} \nonumber \\
&=& \A_e \begin{bmatrix}
N_1^e \\
\vdots \\
N_k^e
\end{bmatrix}^T  \begin{bmatrix} N_1^e,  \cdots, N_k^e
\end{bmatrix} \rho A_e
\label{eq:mass_discretization}
\end{eqnarray}
and the external force potential term $\int_{\Omega} \delta \bm{u} \cdot \bm{b} d\Omega + \int_{\Gamma_{\bm{t}}} \delta \bm{u} \cdot \bar{\bm{t}}d\Gamma$ can be vectorized as follows,
\begin{eqnarray}
\bm{f}_{ext} = \A_e \left[N^e\right] \cdot \bm{b} A_e + \A_e \left[ \hat{N}^e \right] \cdot \bar{\bm{t}} l_e
\end{eqnarray}
where $\left[N^e\right]$ represents the standard elementary shape function matrix, $\left[\hat{N}^e \right]$ represents the degenerated elementary shape function matrix, $l_e$ represents the degenerated line element and $\A_e$ represents the assemble operator of elements.

Therefore, to summarize the formulations above, the motion equation of the Lagrangian system with a cracked region is briefly shown below,
\begin{eqnarray}
\bm{M} \ddot{\bm{u}} + \bm{f}_{int} = \bm{f}_{ext}
\label{eq:discretized_motion_equation}
\end{eqnarray}
An explicit \textit{Newmark} temporal discretizations scheme is utilized in current study, i.e., setting $\beta=0$ in \textit{Newmark} formulae,  so that the time integration scheme is implemented firstly on acceleration at the next time step, i.e.,
\begin{equation}
\begin{aligned}
\bm{M}  \bm{\mathrm{a}}_{n+1} &= \bm{f}^{ext}_{n+1} - \bm{f}^{int} \left( \bm{\mathrm{u}}_{n+1} \right)  \\
&= \bm{f}^{ext}_{\rm n+1} - \bm{f}^{int} \left( \bm{\mathrm{u}}_{n} + \bm{\mathrm{v}}_{n} \Delta t_n + \frac{ \bm{\mathrm{a}}_n \Delta t_n^2 }{2}\right)
\end{aligned}
\end{equation}
Based on diagonal formulation of discretized mass matrix (see Eq.(\ref{eq:mass_discretization})), the nodal acceleration $\bm{\mathrm{a}}_{n+1}$ can be obtained without equation solving. Subsequently, the nodal velocity at next time step $\bm{\mathrm{v}}_{n+1}$ is available through,
\begin{eqnarray}
\bm{\mathrm{v}}_{n+1} = \bm{\mathrm{v}}_n + \left[ \left(1-\gamma \right) \bm{\mathrm{a}}_n + \gamma \rm \bm{\mathrm{a}}_{n+1} \right] \Delta t_n
\end{eqnarray}
and nodal displacement $\bm{\mathrm{u}}_{n+1}$ is updated through without using $\bm{\mathrm{a}}_{n+1}$ 
\begin{eqnarray}
\bm{ \mathrm{u} }_{n+1} = \bm{\mathrm{u}}_n + \bm{\mathrm{v}}_n \Delta t_n + \frac{\bm{\mathrm{a}}_n \Delta t_n^2 }{2} ~.
\end{eqnarray}

\section{A Novel Crack Element Method}
The fracture energy release rate $\mathcal{G}$ plays a pivotal role in the quantitative assessment of crack propagation in structural materials. Technically, it represents the change in potential energy of a system with respect to the crack area and a criterion if the existing crack remains stable or propagates. In linear elastic fracture mechanics, the critical fracture energy release rate $\mathcal{G}_c$ is directly related to the stress density factor $K$ and serves as a material property that defines the toughness of the fracture. Besides, the fracture energy release rate $\mathcal{G}$ is also fundamental to mixed-mode fracture analysis, as it helps distinguish between Mode-I (opening), Mode-II (sliding) and Mode-III (tearing) contributions, guiding the development of mode-dependent failure criteria.

Numerical methods such as the Finite Element Method (FEM), Virtual Crack Closure Technique (VCCT), Cohesive Zone Modeling (CZM) and Extended Finite Element Method (XFEM) are widely employed in fracture mechanics to compute the fracture energy release rate with both flexibility and precision. The Finite Element Method provides a general purpose framework for modeling complex geometries and loading conditions\cite{zienkiewicz2005finite}. Still, it often struggles to resolve stress singularities at crack tips without the use of specialized singular elements. The VCCT, introduced by Rybicki et al.\cite{rybicki1977finite}, efficiently evaluates fracture energy in linear elastic problems using nodal forces and displacements. However, it requires a predefined crack path and fine mesh at the crack front, making it sensitive to mesh size and orientation. Cohesive Zone Modeling, initially proposed by Barenblatt et al.\cite{barenblatt1962mathematical} and Dugdale et al.\cite{dugdale1960yielding}, simulates crack initiation and propagation via incorporating a traction-separation law that captures linear and nonlinear damage evolution. While CZM is effective for modeling ductile and quasi-brittle materials, it demands careful calibration of cohesive parameters. It may encounter convergence difficulties, particularly under dynamic or mixed-mode loading conditions\cite{tvergaard1992relation}. XFEM, developed by Belytschko and co-workers\cite{moes1999finite}, extends the classical FEM by enriching the solution space, allowing cracks to propagate independently of the mesh. This eliminates the need for re-meshing and enables the simulation of arbitrary crack paths. Despite its flexibility, XFEM is mathematically complex, challenging to implement, and can struggle with accurately representing discontinuities in highly nonlinear or heterogeneous materials.

\begin{figure*}[!t]
\centering
\subfloat[]{\includegraphics[width=2.0in]{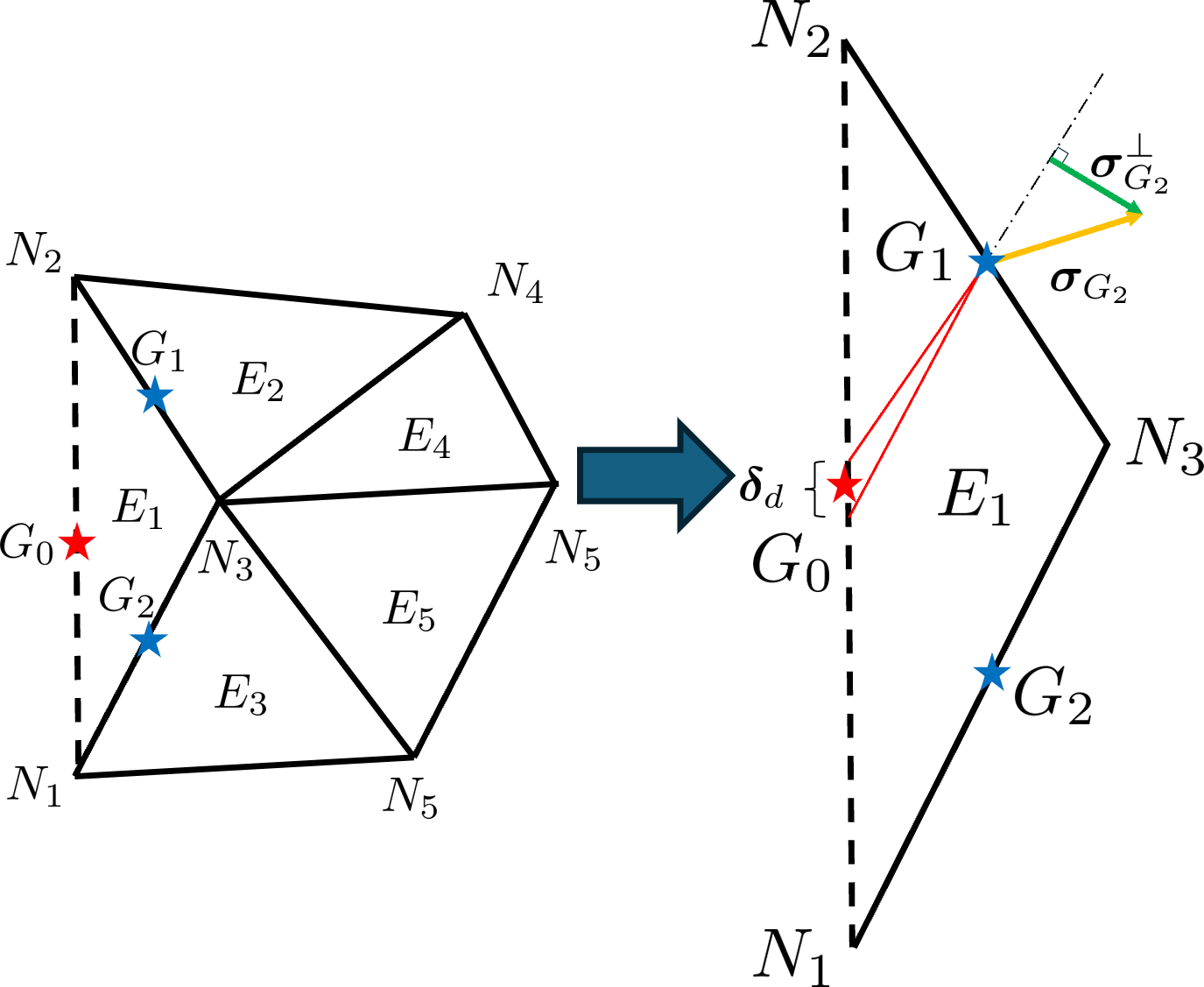}}
\hfil
\subfloat[]{\includegraphics[width=2.0in]{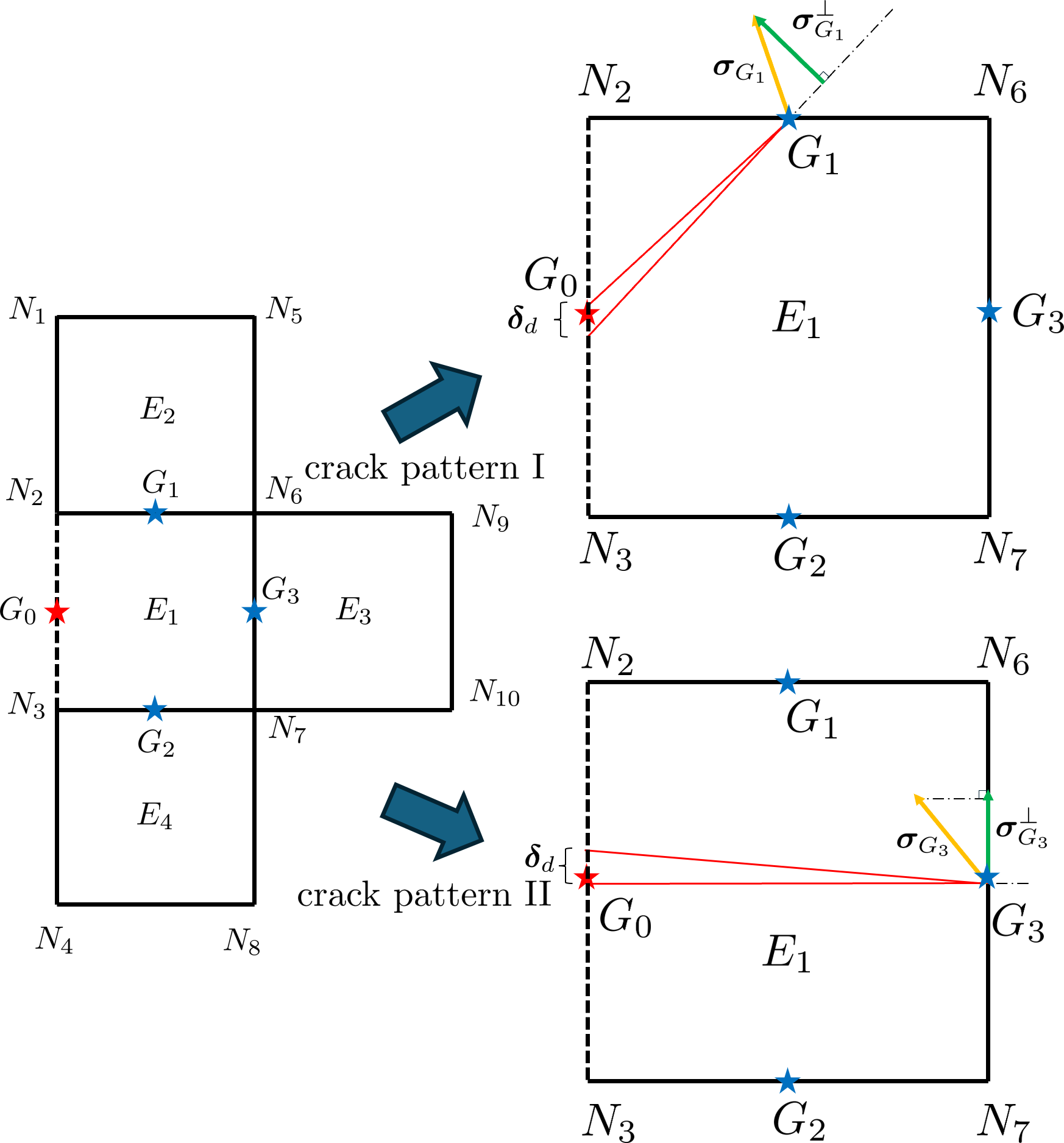}}
   \caption{Element-wise crack type: (a). crack type of Constant Strain Triangle element; (b). crack type of bilinear Quadrilateral element, in which $G_i$ represents edge quadrature point ID, $E_i$ represent element ID, $\bm{\sigma_{G_i}}$ represents the stress state at the location of edge quadrature $G_i$ and $\bm{\sigma}^{\perp}_{G_i}$ represents the normal projection of stress state $\bm{\sigma}_{G_i}$. }
   \label{fig23:Gc_compute_CST_QUAD}
\end{figure*}

While these methods provide powerful tools for fracture analysis, their implementation can be complex, particularly in three-dimensional applications, and they often demand significant computational resources, limiting their practicality for large-scale or real-time simulations. To address these limitations, the Crack Element Method (CEM) is proposed as a robust approach for tracking crack propagation in fracturing solids. This method retains the Finite Element Method as its core framework, introducing a novel element-splitting algorithm and a new formulation of fracture energy release rate to handle crack tip behavior associated with stress singularities accurately. Given the widespread use of Constant Strain Triangle (CST) and bilinear Quadrilateral (Q4) elements in industrial applications, these two element types are selected to illustrate the implementation of the element-splitting algorithm in the proposed CEM approach (see Figure.\ref{fig23:Gc_compute_CST_QUAD}).

In the CST element type, as shown in Figure.\ref{fig23:Gc_compute_CST_QUAD}(a), $E_i$ represents element ID, $G_j$ represents the edge quadrature point ID and $N_k$ denotes node ID. The red star symbol represents the current crack tip at the location of $G_0$, while the two blue stars indicate the candidates for the next crack tip at the locations of $G_1$ and $G_2$, respectively. In the zoomed-in element $E_1$, $\bm{\delta}_d$ represents the stretch of the edge where the crack tip quadrature point is located and is defined as follows,
\begin{eqnarray}
\bm{\delta}_d = \left(  \bm{u}_{N_2} -  \bm{u}_{N_1} \right) \cdot \mathcal{H}\left(\frac{\|  \bm{x}_{N_2} -  \bm{x}_{N_1} \|}{\|  \bm{X}_{N_2} -  \bm{X}_{N_1} \|} - 1 \right) 
\label{eq:delta_d}
\end{eqnarray}
in which $\bm{x}$ and $\bm{X}$ denote deformed and undeformed coordinates of nodes, $\bm{u}$ indicates the displacement of nodes, $\mathcal{H}\left(\cdot \right)$ is the Heaviside function and $\left\| \cdot \right\|$ is the Euclidean norm of a vector. $\bm{\sigma}_{G_2}$ indicates the maximum principal stress at edge quadrature point $G_2$, which is illustrated by a yellow arrow. $\bm{\sigma}_{G_2}^{\perp}$ denotes the normal projection of maximum principal stress at edge quadrature point $G_2$, which is illustrated by a green arrow and is perpendicular to the extended line (shown as a dashed line in Figure.\ref{fig23:Gc_compute_CST_QUAD}(a)) connecting $G_0$ and $G_2$. In bilinear Quadrilateral (QUAD) element type (see Figure.\ref{fig23:Gc_compute_CST_QUAD}(b)), it is important to note that two possible crack patterns distinguish the QUAD element from the CST element in terms of the Crack Element Method: the crack pattern I shows that a potential crack path from edge quadrature point $G_0$ to $G_1$ or $G_2$, that sit at adjacent sides. In contrast, crack pattern II demonstrates that another possible crack path from $G_0$ to to $G_3$ that is located at opposite side. The discrepancy of crack patterns in CST and QUAD elements stems from the geometric differences of element type. Other notations in the QUAD element follow notations in the CST element. 
\begin{figure*}[!t]
\centering
\subfloat[]{\includegraphics[width=2.0in]{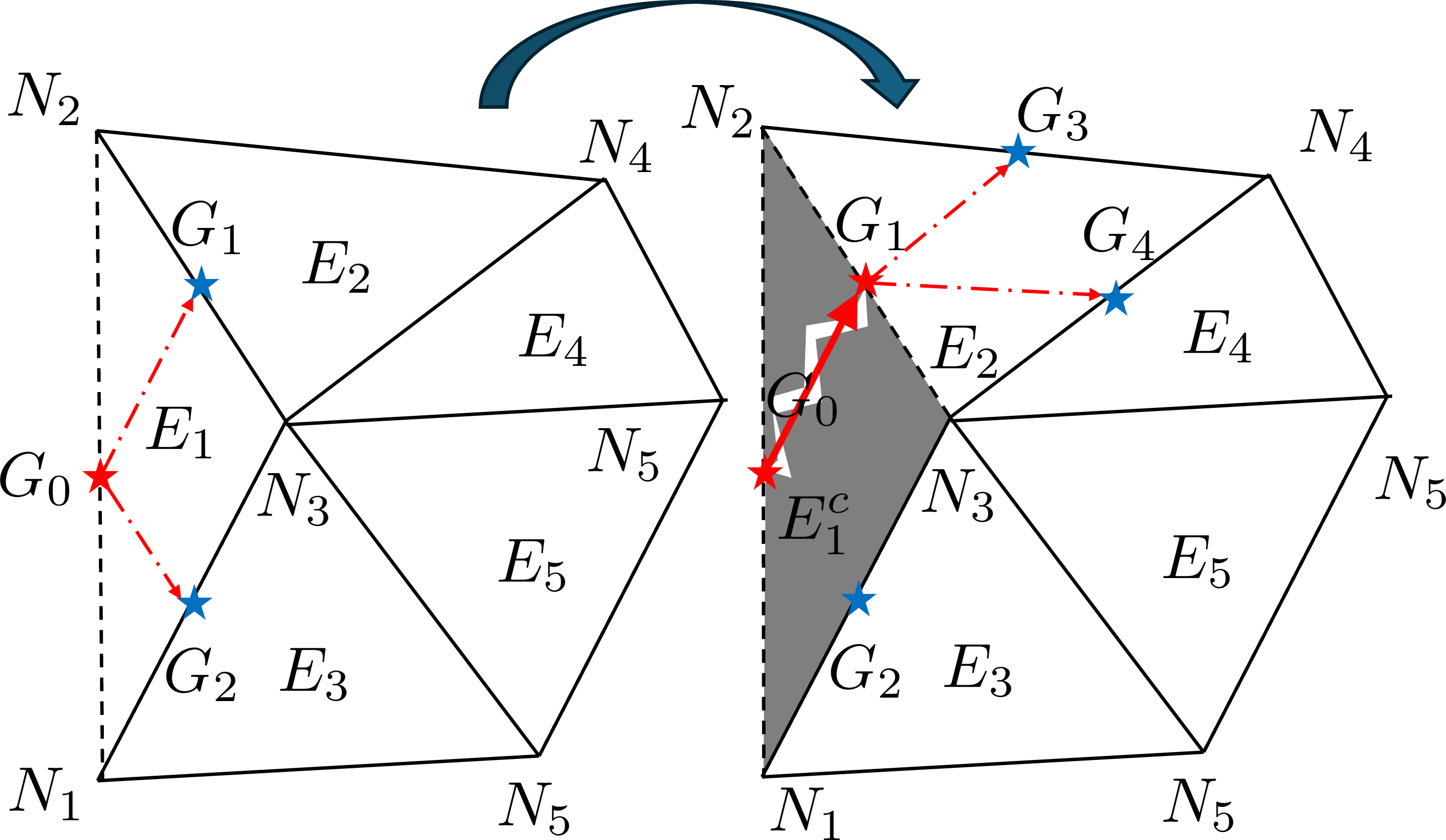}}
\hfil
\subfloat[]{\includegraphics[width=2.0in]{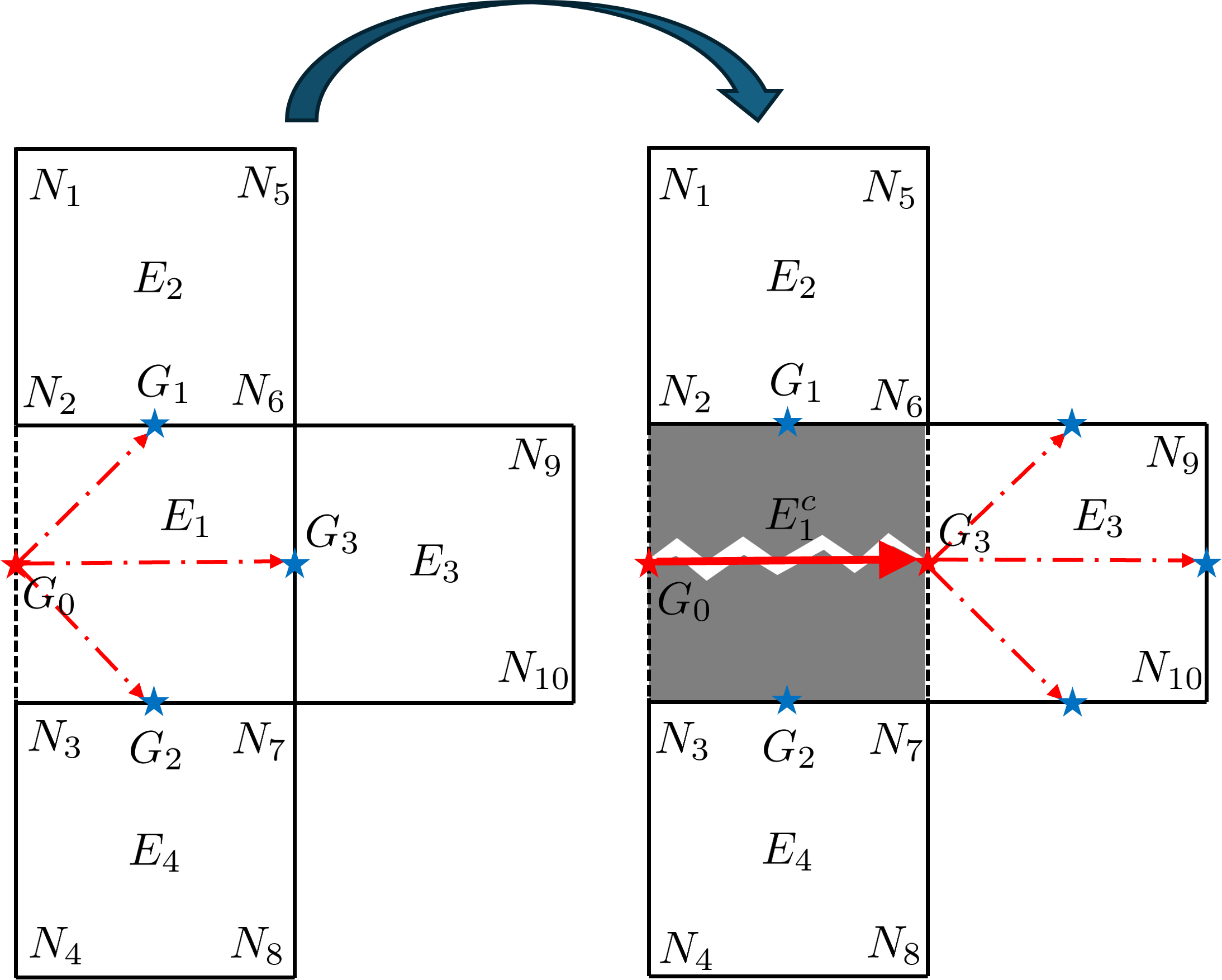}}
\hfil
\subfloat[]{\includegraphics[width=2.0in]{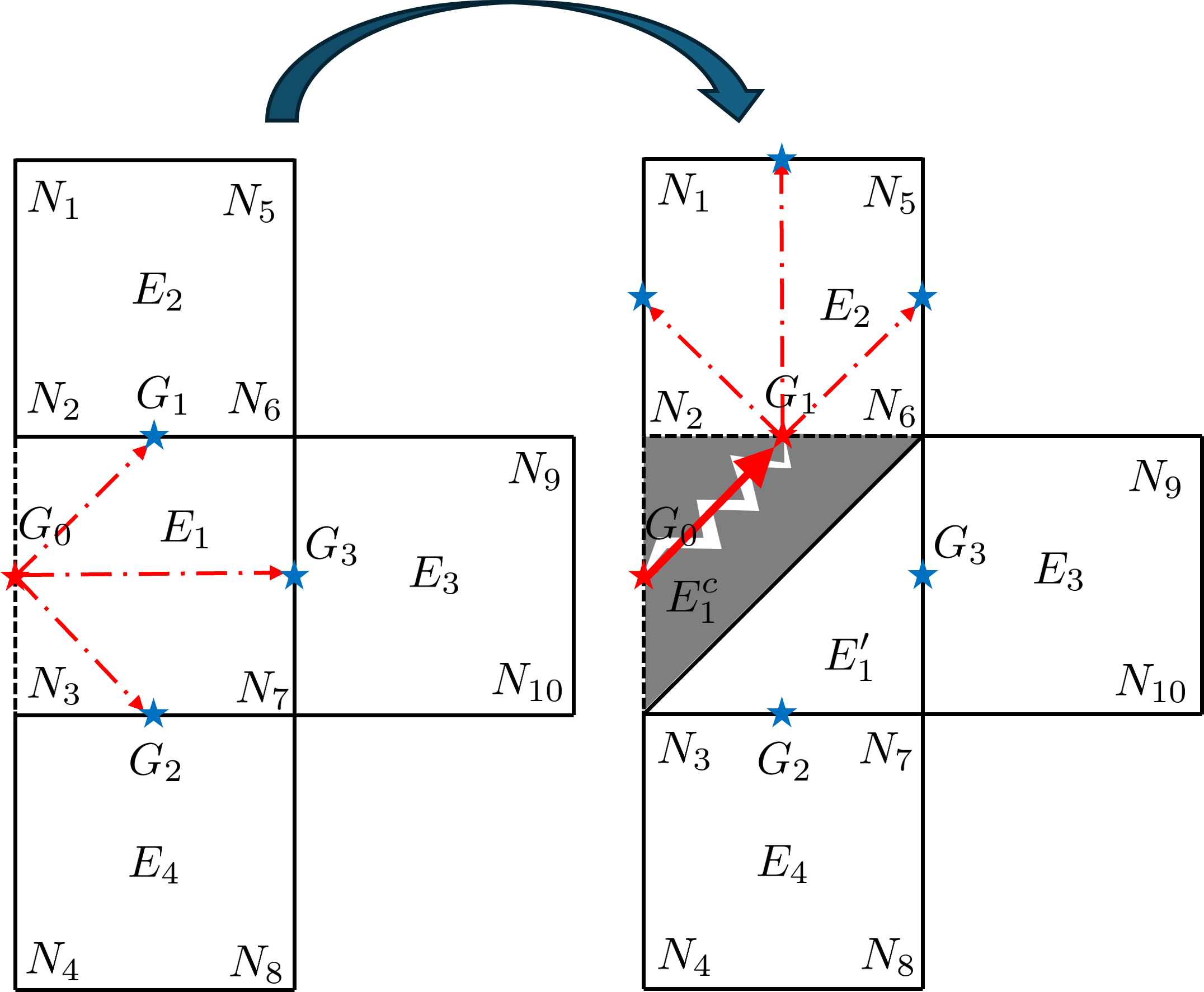}}
   \caption{Intra- and inter-element crack propagation and cracked element reformulation in the CEM: (a). crack propagation in Constant Strain Triangle (CST) element; (b-c). crack propagation in bilinear Quadrilateral (QUAD) element. $N_i$, $G_j$ and $E_k$ denote IDs of node, edge quadrature point and element, respectively. The red star denotes the crack tip currently, the blue stars denote crack tip candidates, the dashed red lines denote possible crack paths and solid red line denotes cracked path within element. The gray portion of element denotes completely failed element, which is skipped in following computation.}
   \label{fig24:crack_prop_CST_QUAD}
\end{figure*}

Based on the proposed element-splitting algorithm, a formulation of fracture energy release rate $\mathcal{G}$ is derived as follows,
\begin{eqnarray}
\mathcal{G}_{G_0} = \frac{\bm{\sigma}_{G_i}^{\perp} \cdot \bm{\delta}_d}{2}
\label{eq:fracture_energy_release_rate}
\end{eqnarray}
when $\mathcal{G}_{G_0} > \mathcal{G}_c$, the crack path forms from current quadrature point $G_0$ to candidate quadrature point $G_i$, in which $\mathcal{G}_c$ is the material critical fracture energy release rate (the subscript $\Box_{c}$ denotes "\textit{critical}").\\
In Constant Strain Triangle element (see Figure.\ref{fig24:crack_prop_CST_QUAD}(a)), current crack tip sits at location of edge quadrature point $G_0$. There are two possible crack paths for it, which are illustrated as red dashed lines and lead to two quadrature points $G_1$ and $G_2$ of adjacent sides. So that two fracture energy release rates are obtained as follows,
\begin{eqnarray}
\mathcal{G}_{G_1-G_0} = \frac{\bm{\delta}_d \cdot \bm{\sigma}_{G_1}^{\perp}}{2}, \qquad \mathcal{G}_{G_2-G_0} = \frac{\bm{\delta}_d \cdot \bm{\sigma}_{G_2}^{\perp}}{2} ~.
\end{eqnarray}
It is assumed here $\mathcal{G}_{G_1-G_0} > \mathcal{G}_{G_2 - G_0} > \mathcal{G}_c$ so crack path forms with moving crack tip from $G_0$ to $G_1$ and element $E_1^c$ (the superscript $\Box^{c}$ denotes the element is completely cracked) is completely failed, as shown in Figure.\ref{fig24:crack_prop_CST_QUAD}(a).\\
In bilinear Quadrilateral element (see Figure.\ref{fig24:crack_prop_CST_QUAD}(b-c)), current crack tip sits at the location of edge quadrature point $G_0$. There are three possible crack paths so that three fracture energy release rates are listed below,
\begin{eqnarray}
&&\mathcal{G}_{G_1-G_0} = \frac{\bm{\delta}_d \cdot \bm{\sigma}_{G_1}^{\perp}}{2}, \nonumber \\
&&\mathcal{G}_{G_2-G_0} = \frac{\bm{\delta}_d \cdot \bm{\sigma}_{G_2}^{\perp}}{2}, \\
&&\mathcal{G}_{G_3-G_0} = \frac{\bm{\delta}_d \cdot \bm{\sigma}_{G_3}^{\perp}}{2} ~. \nonumber
\end{eqnarray}\\
If the crack path to quadrature point $G_3$ forms, the crack tip moves from $G_0$ to $G_3$ and element $E_1^c$ completely fails (see Figure.\ref{fig24:crack_prop_CST_QUAD}(b)). It is worthy noting that if the crack tip chooses to $G_1$ or $G_2$ as the next starting point, element $E_1$ is partially failed rather than completely failed (see Figure.\ref{fig24:crack_prop_CST_QUAD}(c)). In other words, if the crack path forms from $G_0$ to $G_1$, only the upper triangle (grey portion) is fractured while the lower triangle $E_1^{\prime}$ still works (if from from $G_0$ to $G_2$, then vice versa.). Therefore, the QUAD element formulation is degenerated into the CST element formulation in the following internal/external force computation. 

\section{Numerical Benchmark Examples}
Three classical transient dynamic crack propagation problems in quasi-brittle materials are presented to demonstrate the effectiveness of the proposed method. In addition, two case studies involving mechanically and thermally induced crack propagation in multilevel interconnects are included to highlight the method’s practical relevance to the semiconductor application. For consistency, all simulations are performed using the constant strain triangle element formulation.

\subsection{Kalthoff-Winkler experiment}
\begin{figure}[!t]
\centering
\includegraphics[width=3.0in]{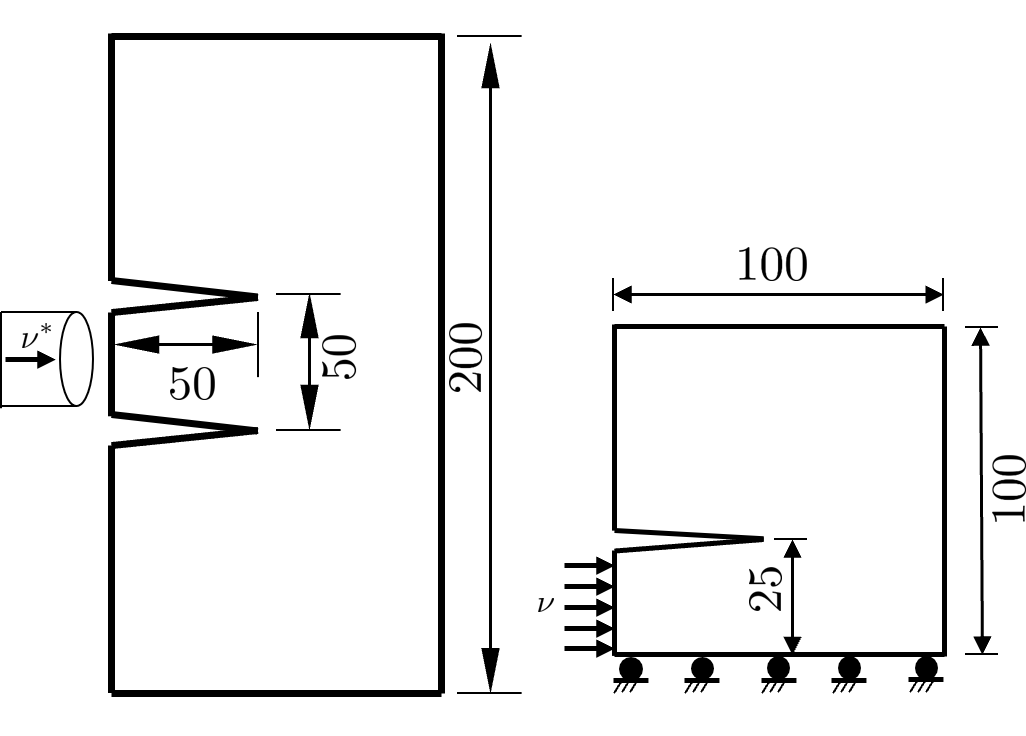}
    \caption{The whole geometry, boundary condition of Kalthoff-Winkler plate experiment and upper half of the Kalthoff-Winkler plate.}
    \label{fig1: kalthoff-plate}
\end{figure}

The shear impact loading experiments by Kalthoff et al.\cite{kalthoff1988failure,kalthoff2000modes} are famous as a verification of transient-dynamic crack simulation in quasi-brittle materials. In Figure.\ref{fig1: kalthoff-plate}, an external transient loading is applied on the left center of the plate at various velocities. Since our current numerical implementation only considers brittle materials, the impact velocity is restricted to a value of $\nu^{\ast} = 33m/s$, allowing the steel plate to exhibit brittle fracture mode. According to \cite{kalthoff1988failure} and \cite{kalthoff2000modes}, ramping applied velocities on a steel plate can generate a series of transition responses from brittleness to ductility. Experiments show that the crack will initiate at the pre-crack tip and propagate along a $70^{\circ}$ angle from the horizontal line. The experiment is considered as a classic transient dynamic brittle fracture benchmark, which has been extensively studied by many crack numerical methods\cite{rabczuk2008discontinuous,song2009cracking,borden2012phase}.

The material properties in simulation are shown as below: $E=190\ GPa$, $\nu = 0.3$, $G_f = 2.213 \times 10^4\ J/m^2$, $\rho = 8000\ kg/m^3$. Besides, the dilatational, shear and Rayleigh wave velocities are $v_d = 5654\ m/s$, $v_s=3022\ m/s$, $v_R = 2803\ m/s$. The Dirichlet boundary conditions are: external impact is horizontally applied at the left center part with the specified velocity, and the bottom boundary of the upper half plate are constrained vertically, as shown in Figure.\ref{fig1: kalthoff-plate}(b). And the applied velocity is firstly ramp-up and then keeps constant, 
\begin{equation}
v =
\begin{cases} 
\frac{t}{t_0}v_0,  & \text{if } \ t \le t_0, \\
v_0, & \text{if } \ t > t_0.
\end{cases}
\end{equation}
in which, $v_0 = 16.5\ m/s$ and $t_0 = 1\ \mu s$.

\begin{figure}[!t]
\centering
\includegraphics[width=3.5in]{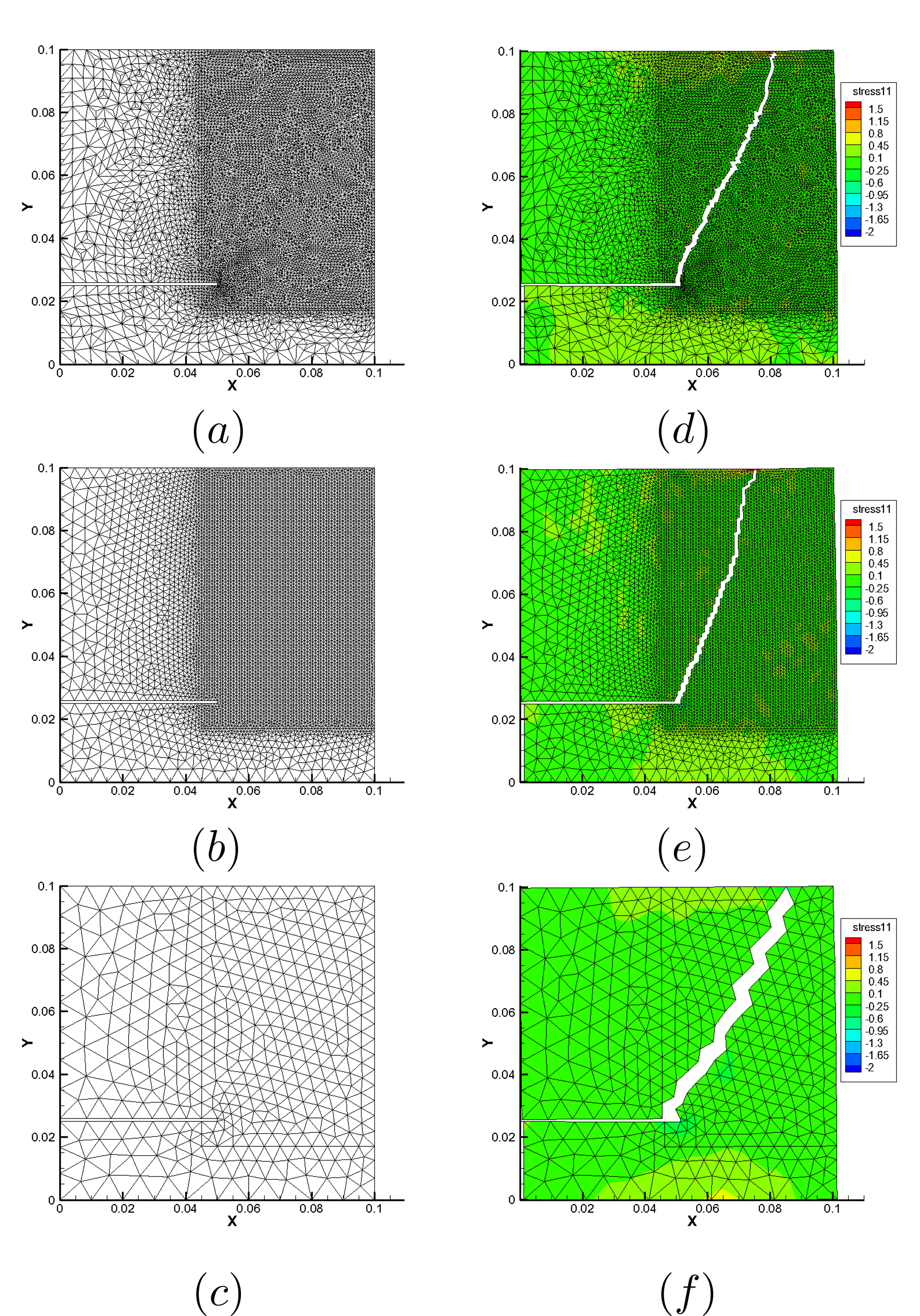}
    \caption{Three representative triangle meshes are illustrated: (a). irregular fine mesh; (b). regular fine mesh; (c). coarse mesh. The final crack patterns of the three representative meshes are illustrated: (d). irregular fine mesh; (e). regular fine mesh; (f). coarse mesh.}
\label{fig2: kalthoff-meshes-crack-pattern}
\end{figure}

Three different representative constant strain triangle meshes are picked up to conduct simulation: the first one has an irregular triangle fine mesh in the crack growth zone with $7212$ nodes and $14158$ elements, the second one has a regular fine mesh in the crack growth zone with $6727$ nodes and $13199$ elements, and the third one is a coarse triangle mesh in the crack growth zone with $434$ nodes and $790$ elements, as shown in Figure.\ref{fig2: kalthoff-meshes-crack-pattern}(a-c).

The crack patterns at the final stage are illustrated in Figure.\ref{fig2: kalthoff-meshes-crack-pattern}(d-f). Compared to numerical reference results as in \cite{borden2012phase} and experimental results in \cite{kalthoff1988failure}, the crack angle in the proposed method is around $65^{\circ}\sim 70^{\circ}$, which is comparable to reference numerical results ($65^{\circ}$) and experimental data ($70^{\circ}$). It is noted that whether an irregular mesh with random edge orientations or a regular mesh with fixed edge orientations, the crack propagations are stably approaching $70^{\circ}$. More importantly, the proposed method still accurately captures the crack path in the coarse mesh with only $434$ nodes and $790$ elements, greatly outperforms other numerical methods, especially compared to smear crack approaches, which usually demands at least ten thousands of quadrilateral elements\cite{bui2022simulation,borden2012phase,song2008comparative,wu2016meshfree}. 

\begin{figure}[!t]
\centering
\includegraphics[width=3.5in]{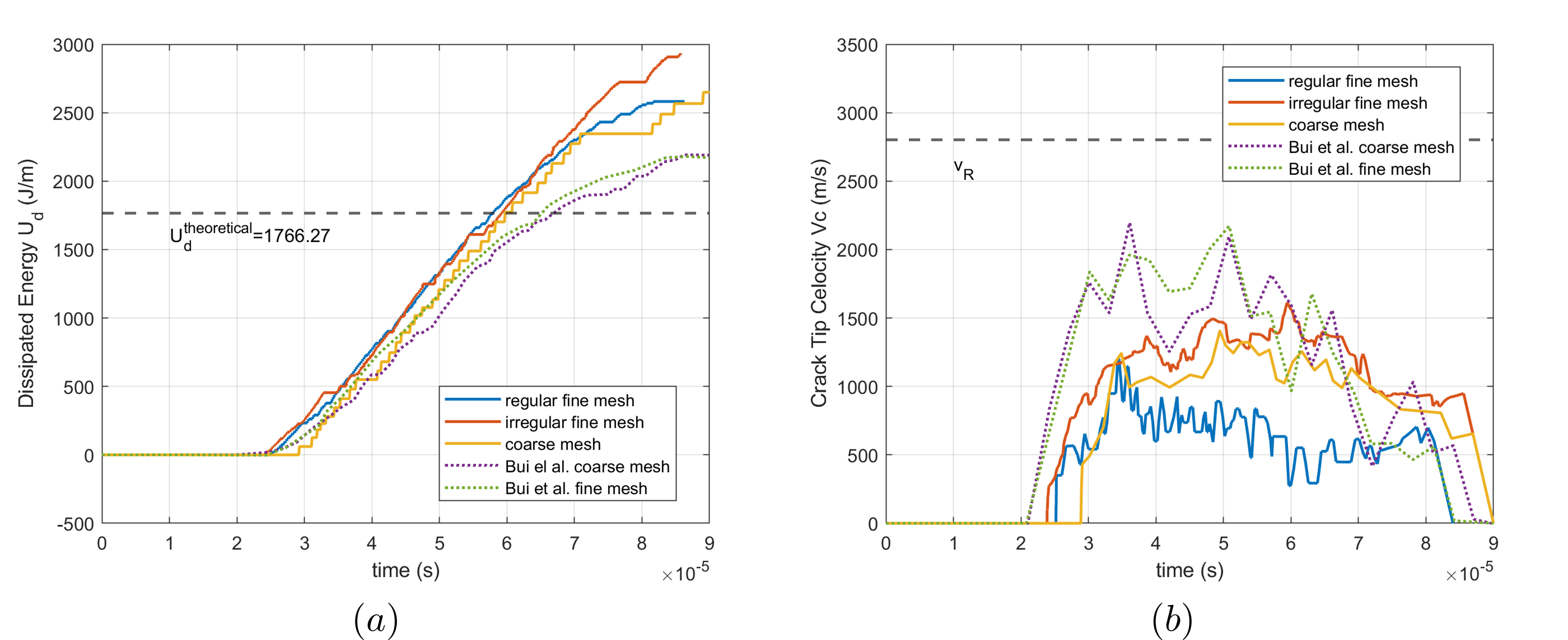}
    \caption{Comparison of (a). dissipated energy and (b). crack tip velocity among three representative grids and two reference grids\cite{bui2022simulation}.}
    \label{fig7: Kalthoff-Ud-cvel}
\end{figure}

The results of the dissipated energy and the velocities of the crack tip in the three representative models are summarized in Figure.\ref{fig7: Kalthoff-Ud-cvel}. In the dissipated energy comparison (Figure.\ref{fig7: Kalthoff-Ud-cvel}(a)), it is found that the dissipated energy across three different meshes is generally consistent, slightly higher than the reference numerical results\cite{bui2022simulation} but still within a reasonable range. The reason why more energy is dissipated in the proposed CEM compared to other numerical results\cite{bui2022simulation} may come from discrepancy in dealing with partially failed element: in CEM, the fracture state of an element is either fully cracked or perfect, while in Bui et al.\cite{bui2022simulation} the damage index $d$ ranges $0 \sim 1$, and only those elements with $d > 0.95$ are categorized as completely failed. The theoretical value $U_d = 1766.27\ J/m$ is calculated based on the fracture energy $G_f = 2.213 \times 10^4 \ J/m^2$ and an ideal straight crack propagation path along $70^{\circ}$ whose entire length is about $79.81\ mm$. The reason that more energy is dissipated in CEM compared to the theoretical value is probably that the theoretical dissipated energy value in the distorted cracking path and partially damaged elements may be overlooked. 

The velocities of crack tip comparison are also studied in Figure.\ref{fig7: Kalthoff-Ud-cvel}(b). It is observed that the time at which crack initiation begins in the three models are basically the same (crack initiates in coarse mesh slightly later than in the two fine meshes). The regular fine mesh may have more fluctuations than the other two. It may be caused by relatively regular crack propagation (see Figure.\ref{fig2: kalthoff-meshes-crack-pattern}, the regular fine mesh crack pattern shows a periodic path, especially at the cracking early stage). Besides, it is found that the crack-tip velocity of fine regular mesh decreases after $35\ \mu s$ while the crack-tip velocities of coarse mesh and irregular mesh increase up to $1600\ m/s$.  The reason may be due to a different crack path: regular fine mesh can have a straighter crack path compared to the other two meshes, which leads to lower crack-tip velocities.

\subsection{Three-point Support Bending Beam}
\begin{figure}[!t]
\centering
\includegraphics[width=3.0in]{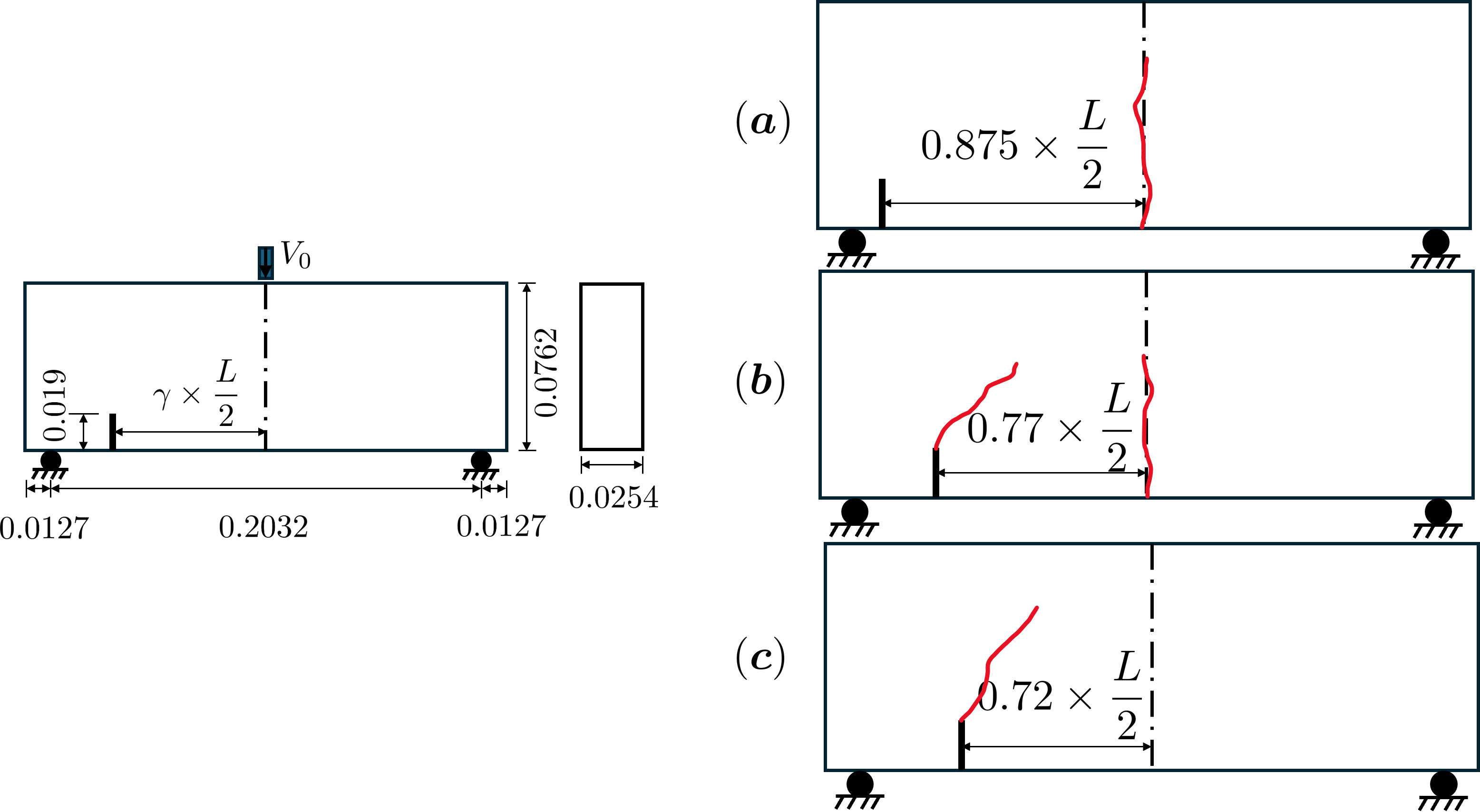}
    \caption{The three-point bending beam with a pre-notch for a mixed mode crack benchmark \cite{john1990mixed}: (a). mode-I crack pattern at midspan of beam; (b). both mixed mode and mode-I crack pattern; (c). mixed mode crack pattern.}
    \label{fig8: bbeam-concept-experiment}
\end{figure}

John et al. conducted a concrete beam experiment under impact loading to study mixed-mode dynamic crack propagation\cite{john1990mixed}. According to the reported experimental results, the location of pre-notch determines crack propagation patterns in the concrete beam (see Figure.\ref{fig8: bbeam-concept-experiment}). In the experiment, an offset parameter $\gamma$ is used to characterize the pre-notch location at the beam:
\begin{equation}
d_{pre-notch} = \gamma \times \frac{L}{2}
\end{equation}
in which, $d_{pre-notch}$ represents the distance from the pre-notch to the midspan of the beam, and $L$ is the distance between the two supports. With increasing value of $\gamma$ from $0$ to $0.9$, the pre-notch is approaching from the midspan of the beam to the left end support. Meanwhile, the crack propagation pattern gradually shifts from pure mode-I fracture to a combination of mode-I and mixed mode fracture, finally resulting in a mixed mode fracture pattern (see Figure.\ref{fig8: bbeam-concept-experiment}(b-d)).

The proposed CEM is applied to simulate the gradual evolution of crack patterns with different distance parameters $\gamma$. In the finite element model, a relatively coarse grid, specifically, $30 \times 91$ mesh, is employed to discretize the beam, which measures $0.2286\ m \times 0.0762\ m$. A material fracture strength-dependent criterion is used to initiate a crack at midspan, where no pre-notch is present to concentrate stress. The material properties of the concrete are provided: $E=28\ GPa$,\ $\nu = 0.2,\ G_f=22\ J/m^2,\ \rho=2400\ kg/m^3$ and fracture strength $f_t = 8\ MPa$. The impact loading in the experiment can be characterized as a ramping load as follows:
\begin{equation}
v =
\begin{cases} 
\frac{t}{t_0}v_0,  & \text{if }\ t \le t_0, \\
v_0, & \text{if } \ t > t_0.
\end{cases}
\end{equation}
in which, $t_0 = 196\ \mu s$ and $v_0 = 60\ mm/s$.
\begin{figure}[!t]
\centering
\includegraphics[width=3.0in]{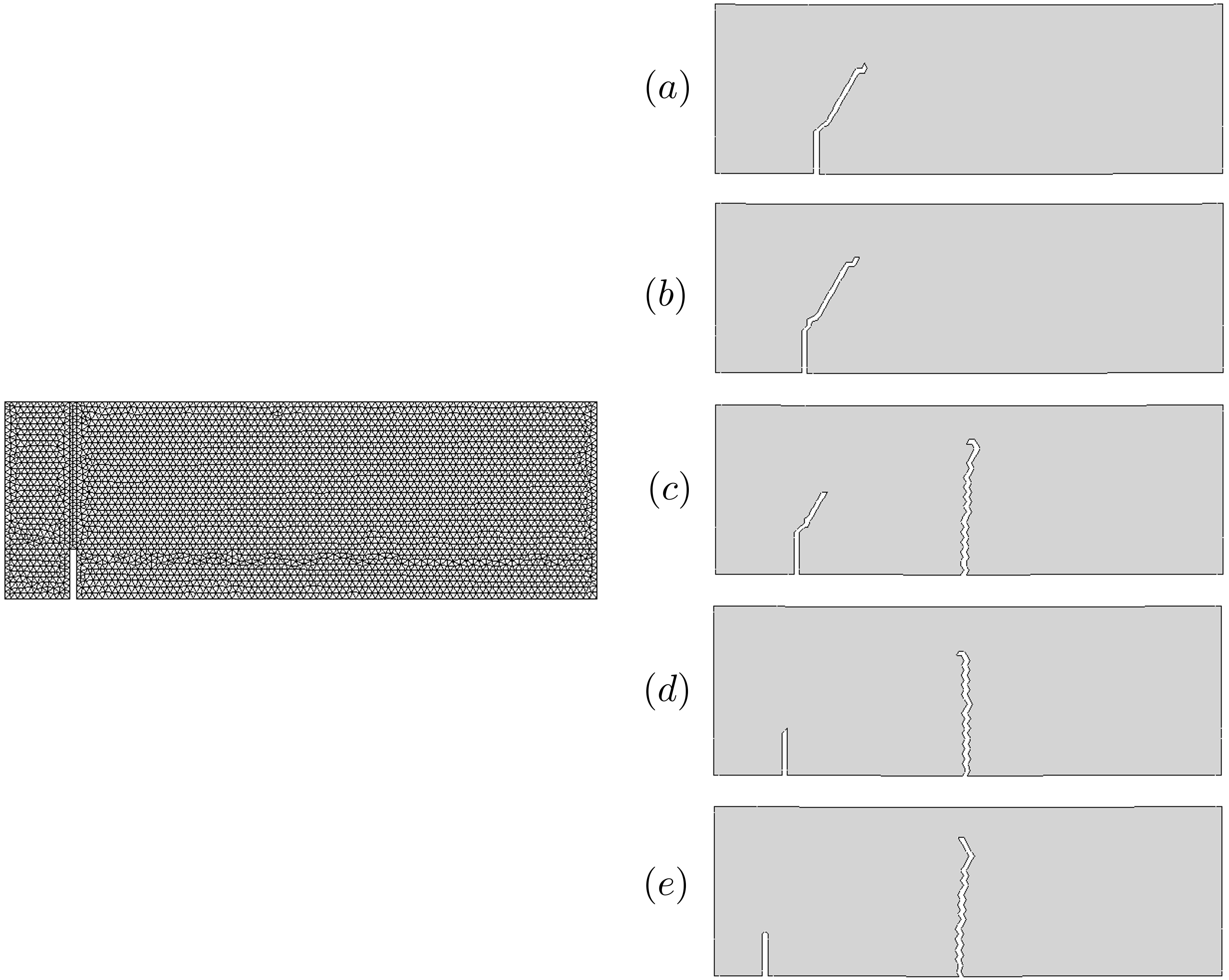}
    \caption{Grids of three-point support bending beam model and crack patterns for different location parameters $\gamma$: (a). location parameter $\gamma = 0.675$; (b). location parameter $\gamma = 0.73$; (c). location parameter $\gamma = 0.765$; (d). location parameter $\gamma = 0.81$; (e). location parameter $\gamma = 0.9$. }
    \label{fig9: bbeam-numerics}
\end{figure}

A series of numerical examples with different location parameters $\gamma$ for the pre-notch are conducted, including $\gamma = 0.675,\ 0.73,\ 0.765,\ 0.81,\ 0.9$ (see Figure.\ref{fig9: bbeam-numerics}(b-f)). The same finite element grid $30 \times 91$ is used for the five cases (see Figure.\ref{fig9: bbeam-numerics}(a)). The time step size is also fixed at $\Delta t=0.1\ \mu s$. The various crack propagation patterns corresponding to different values of parameters $\gamma$ are shown in Figure\ref{fig9: bbeam-numerics}(b-f). It is found that the critical value of the location parameter $\gamma$, which determines the appearance of both mode-I crack and mixed mode crack pattern is around $0.765$. The value $\gamma > 0.765$ results in pure mode-I crack patterns (Figure.\ref{fig9: bbeam-numerics}(e-f)) while the value $\gamma < 0.765$ results in mixed mode crack patterns (Figure.\ref{fig9: bbeam-numerics}(b-c)). 

In Table-\ref{tbl-1:T3beam}, the determined critical location parameter $\gamma_t$ from various methods and experiments is reported. It is found that the critical location parameter $\gamma_t$ from the proposed CEM is consistent with the experimental result compared to other numerical results. It is noted that the studies done in \cite{belytschko2000element,zi2005extended} artificially introduced a non-trivial pre-notch at the mid-span of the beam while the crack initiates naturally in the proposed CEM. 
\begin{table}[!t]
\caption{The comparison of critical location parameter $\gamma_t$ among experiment, the present research and other numerical methods\label{tbl-1:T3beam}}
\centering
\begin{tabular}{|c||c|}
\hline
Approaches & $\gamma_t$\\
\hline
Experiment\cite{john1990mixed} & $0.77$\\
3D cohesive model\cite{ruiz2001three} & $0.6$  \\
XFEM\cite{zi2005extended} & $0.635$  \\
Element free Galerkin method\cite{belytschko2000element} & $0.734$  \\
The Crack Element Method  & $0.765$  \\
\hline
\end{tabular}
\end{table}

\begin{figure}[!t]
\centering
\includegraphics[width=3.5in]{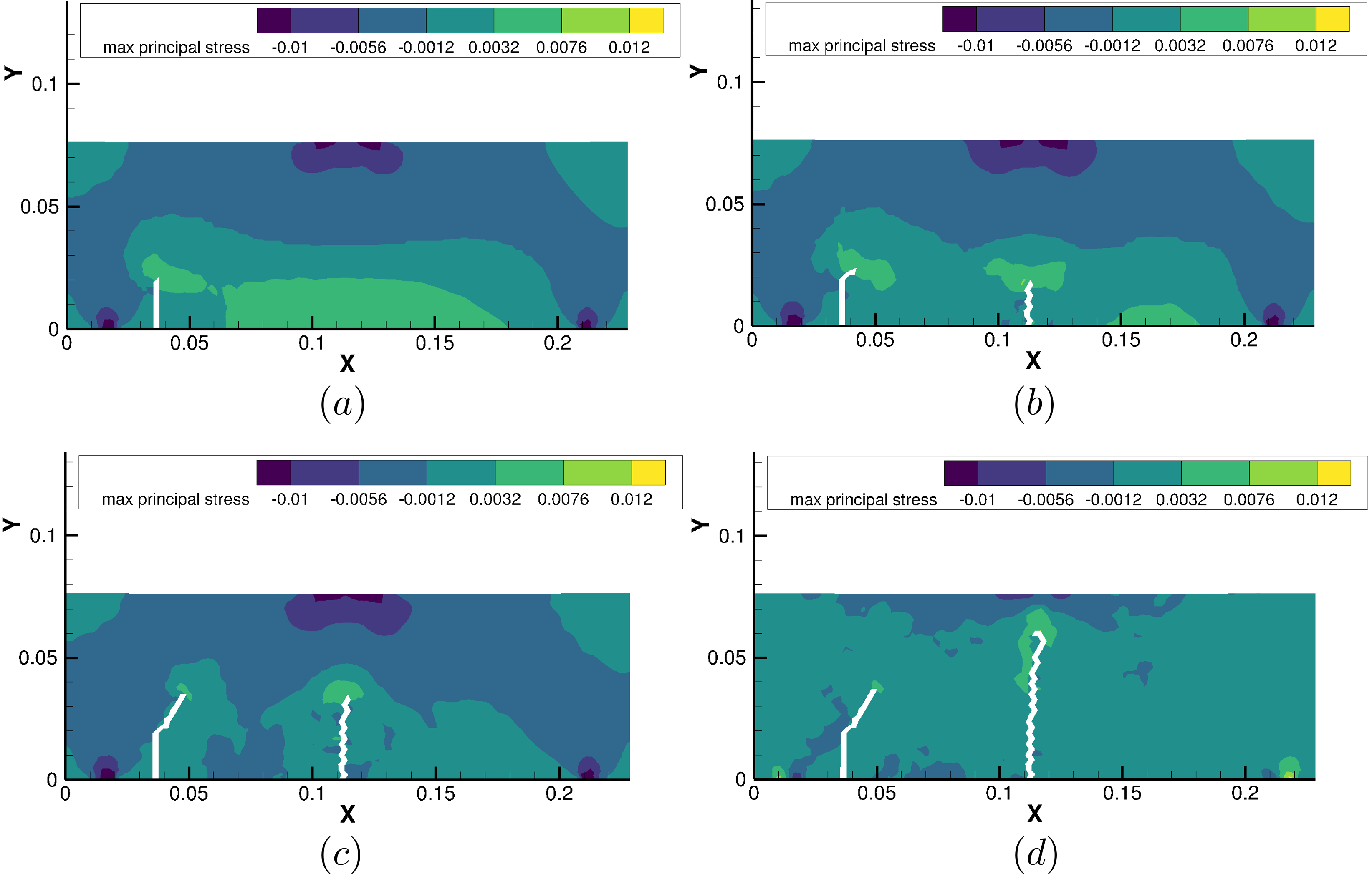}
    \caption{Animation of crack path progressing at the case of $\gamma_t = 0.765$: (a). $t = 1.0985 \times 10^{-3}\  s$; (b).  $t = 1.1674\times 10^{-3} s$; (c). $t = 1.1829\times 10^{-3} s$; (d). $t = 1.2438\times 10^{-3} s$. }
    \label{fig10: bbeam-crack-anima}
\end{figure}

In addition, the progression of the crack path with $\gamma_t = 0.765$ is captured in Figure.\ref{fig10: bbeam-crack-anima}. The whole process is as follows: at time $t = 1.0985 \times 10^{-3}\  s$, the crack initiates at the pre-notch tip. With increasing external loads (such as at time $t = 1.1674\times 10^{-3} s$ and $t = 1.1829\times 10^{-3} s$), the cracks at the midspan of the beam and the pre-notch initiate and continue to grow. At time $t = 1.2438\times 10^{-3} s$, the crack path at the pre-notch basically stops while the crack at the mid-span of beam continues growing to the location where the loads are applied. 

\subsection{Compact Compression Specimen}
This numerical example demonstrates the highly accurate and sensitive capability of capturing the crack tip in dynamic simulation: even for curved and arbitrary paths, can still be traced easily. Rittel et al used PMMA (Polymethyl methacrylate) specimen to complete a series of experiments.\cite{rittel1996investigation}. Following the numerical simulation, many researchers studied this experiment by their proposed modeling methods, including cohesive models\cite{paulino2010adaptive}, Extended FEM\cite{asareh2020general,menouillard2008mass}.

\begin{figure}[!t]
\centering
\includegraphics[width=3.0in]{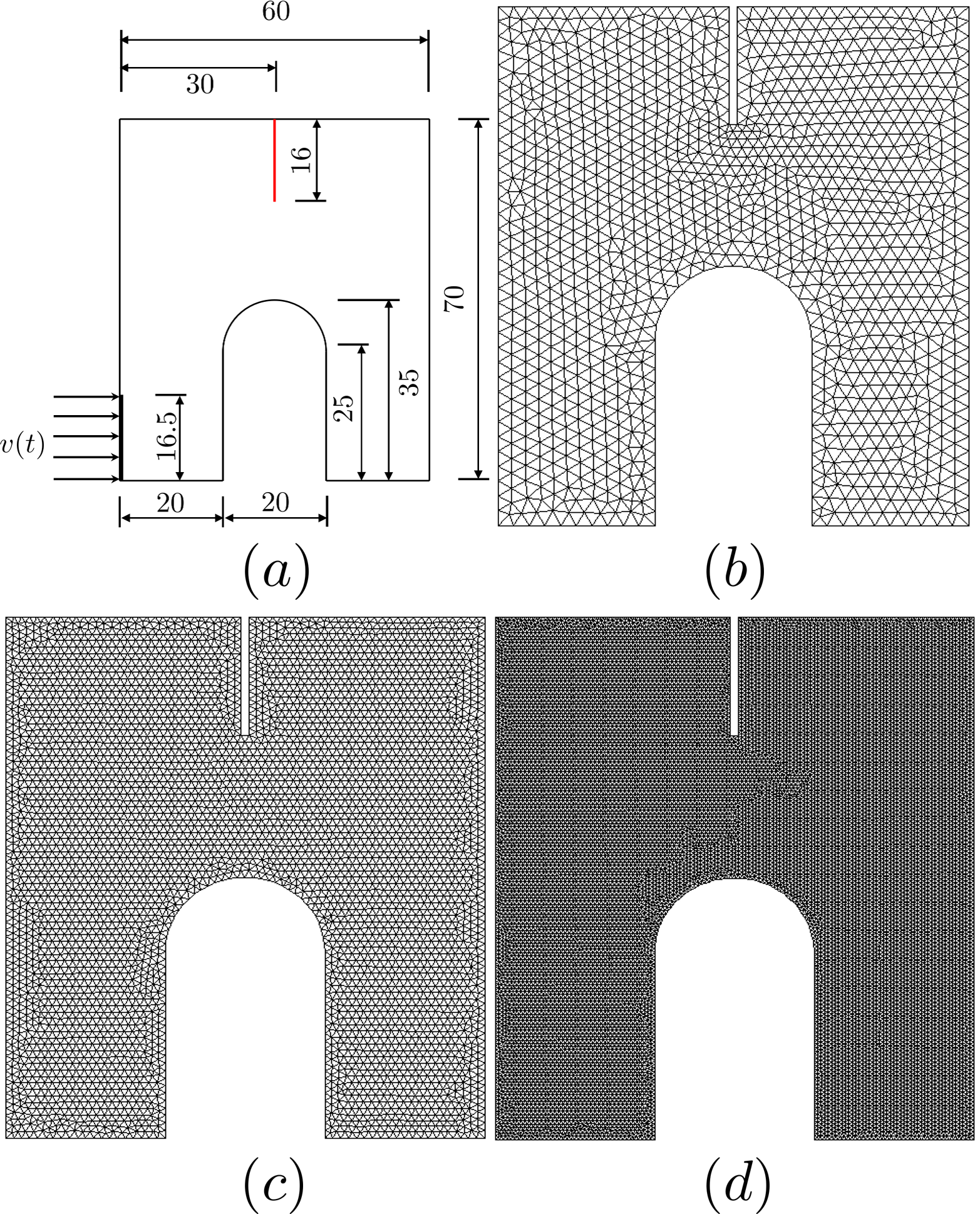}
    \caption{(a). The model geometric dimensions (unit: mm) and boundary conditions of the compact compression experiment. The pre-notch is illustrated as a red line. (b). The grid with 1287 Nodes and 2393 Elements; (c). The grid with 4384 Nodes and 8431 Elements; (d). The grid with 16753 Nodes and 32797 Elements.}
    \label{fig17: ccompression-model-grids}
\end{figure}

The model's geometric dimensions are provided in Figure.\ref{fig17: ccompression-model-grids}(a). The impact by Hopkinson bar in the experiment is located at left bottom left and is modeled by the prescribed velocity as 
\begin{equation}
v =
\begin{cases} 
\frac{t}{t_0}v_0,  & \text{if }\ t \le t_0, \\
v_0, & \text{if } \ t > t_0.
\end{cases}
\end{equation}
in which, $v_0 = 20\ m/s$ and $t_0 = \ 40 \mu s$. Other boundary conditions of the model are unspecified. 

The material properties utilized in the simulation are as follows\cite{asareh2020general}: Young's modulus $E = 5.76 \ GPa$, Poisson ration $\nu = 0.42$, fracture energy release rate $G_f = 352.3\ J/m^2$, density $\rho_0 = 1180\ kg/m^3$, tensile strength $\sigma_c = 129.6 \ MPa$ and Rayleigh wave speed based on material constants given above is $v_R = 1237.5\ m/s$. The total time of the experiment process in simulation is $T = 140\times 10^{-6} s$ and a fixed time step $\Delta t = 1.0 \times 10^{-8} s$. Three different grids, including 1287 Nodes with 2393 Elements, 4384 Nodes with 8431 Elements, and 16753 Nodes with 32797 Elements (see Figure.\ref{fig17: ccompression-model-grids}(b-d)), are generated to validate the robustness of the proposed method. 

\begin{figure}[!t]
\centering
\includegraphics[width=3.0in]{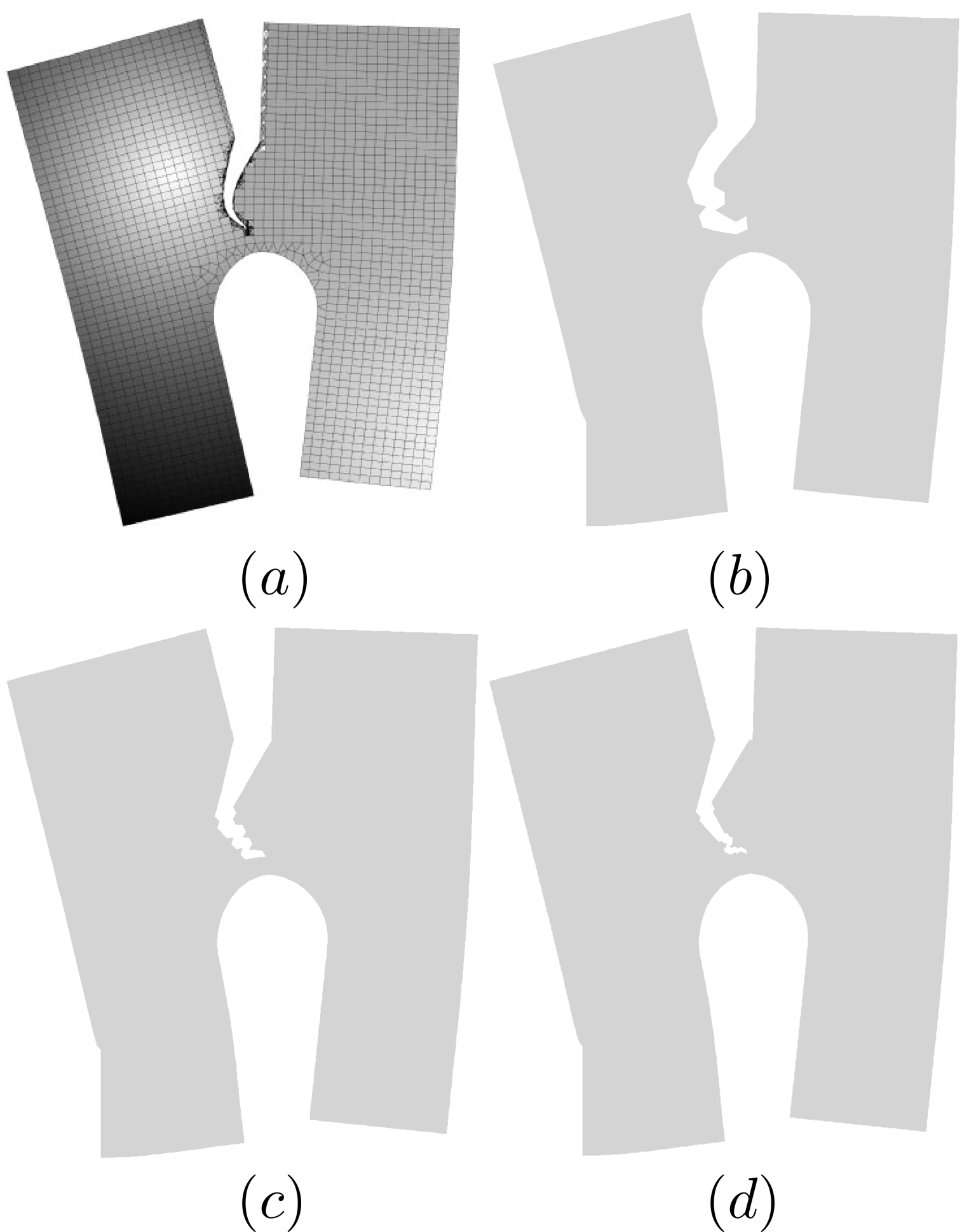}
    \caption{(a). Reference crack path\cite{menouillard2006efficient}; (b). The crack path with 1287 Nodes and 2393 Elements; (c). The crack path with 4384 Nodes and 8431 Elements; (d). The crack path with 16753 Nodes and 32797 Elements.}
    \label{fig18: ccompression-crack-paths}
\end{figure}

The crack propagation paths predicted by the three different grids using the proposed method are shown in Figure.\ref{fig18: ccompression-crack-paths}. Compared qualitatively with crack patterns in the reference result\cite{menouillard2006efficient}, the predicted curved crack path in an arc shape is accurately captured by the CEM leading to the conclusion that the proposed crack tracking algorithm can accurately capture crack growth even under very complicated stress conditions. 

\begin{figure}[!t]
\centering
\includegraphics[width=3.5in]{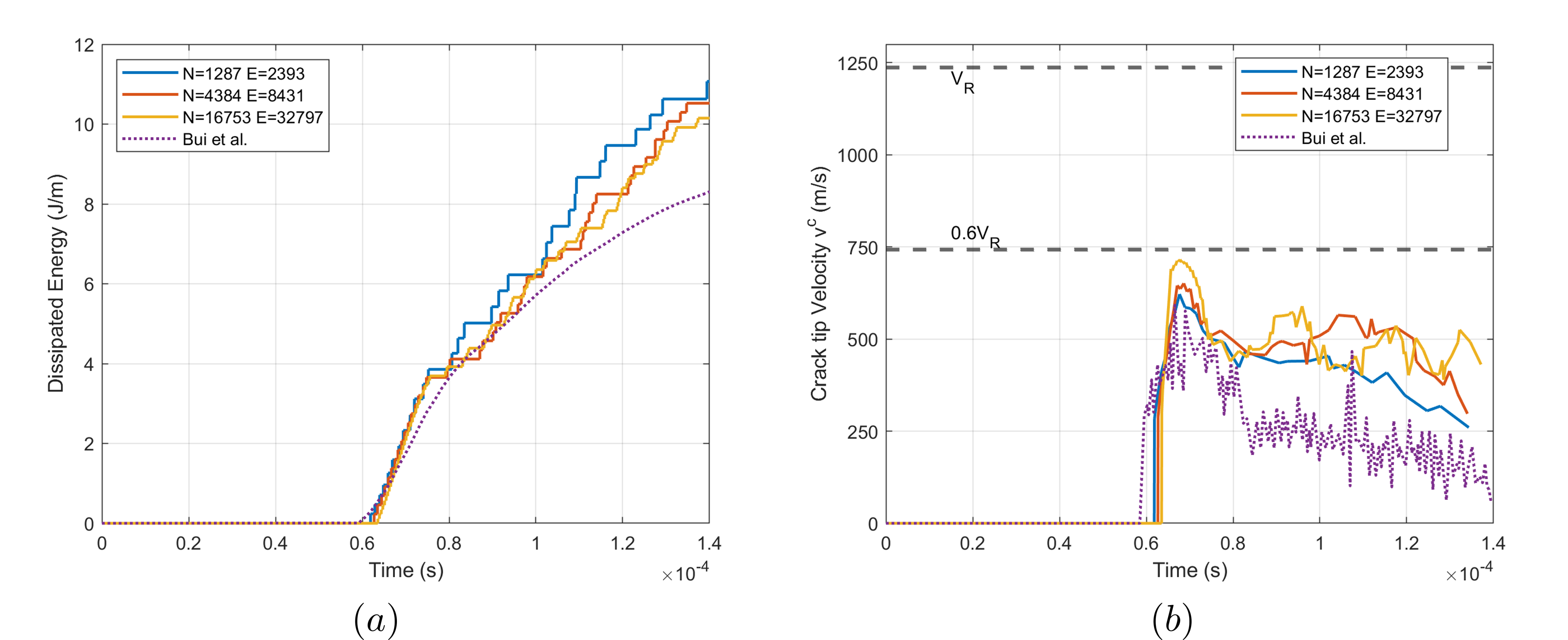}
    \caption{(a). Dissipated energy evolution during crack propagation; (b). crack tip velocity evolution during crack propagation (reference results from Bui et al.\cite{tran2024nonlocal}).}
    \label{fig19: ccompression-energy-cvel}
\end{figure}

Two quantitative comparison results, including dissipated energy and crack tip velocity, are provided in Figure.\ref{fig19: ccompression-energy-cvel}. It is shown that the crack initiates at $61\mu s$. The crack tip velocity increases to about $400\ m/s$, then rises to a peak of $700\ m/s$ at around $70\ \mu s$ and decreases to about $500\ m/s$ after $80\mu s$. In the whole loading time, the crack tip velocity is constrained to $0.6 v_R = 750\ m/s$, i.e., $60 \%$ of the Rayleigh wave speed.  

\begin{figure}[!t]
\centering
\includegraphics[width=3.5in]{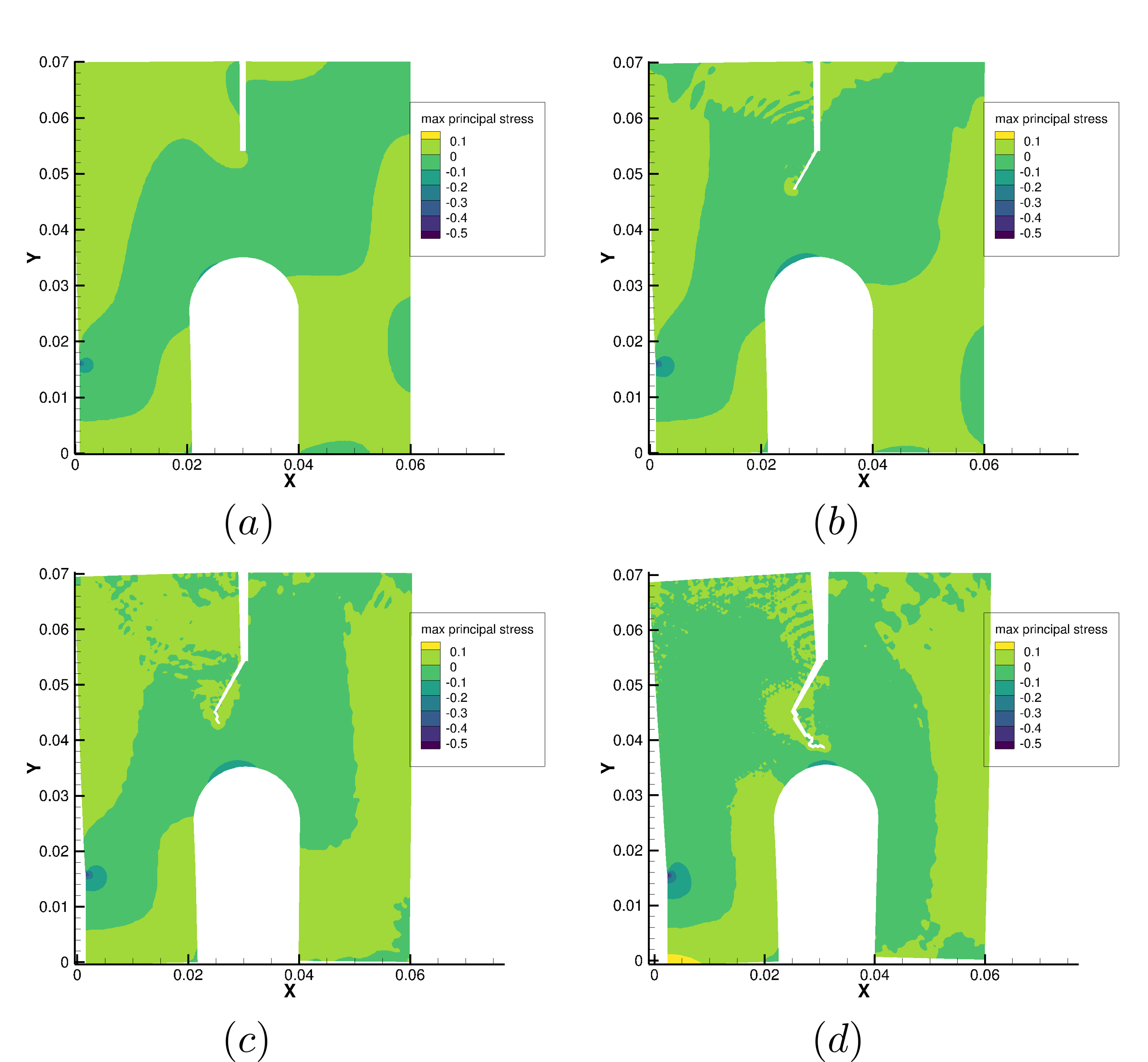}
    \caption{Crack path evolution and maximum principal stress contour with $16753$ nodes and $32797$ elements model: (a). time $t=61.6\ \mu s$; (b). time $t=75.6\ \mu s$; (c). time $t=98\ \mu s$; (d). time $t=140\ \mu s$.}
    \label{fig20: ccompression-crack-evolution}
\end{figure}

Furthermore, the crack propagation with $16753$ nodes and $32797$ elements model is illustrated as in Figure.\ref{fig20: ccompression-crack-evolution}. The crack initiates at the notch tip and curves as it extends toward the left side of the specimen. Ultimately, the crack follows a path resembling an arc. Despite the specimen’s geometric symmetry, the asymmetry in the applied loading causes uneven deformation. This, in turn, leads to an irregular damage distribution that the CEM successfully captures.

\subsection{Cracking in Cu/Ultra Low-k Multilevel Interconnects}
Chip Package Interaction (CPI) and the associated thermomechanical stress in microchips pose significant reliability concerns, particularly in on-chip interconnect stacks. These stresses arise due to mismatches in the thermal expansion coefficients of different materials within the package, resulting in mechanical deformations during thermal cycling. As a result, the integrity of the interconnect structures is compromised, increasing the likelihood of failure. Two primary failure mechanisms are commonly observed: adhesive failure, which occurs due to delamination along the Cu/dielectric interfaces, and cohesive failure, which involves the fracture of the dielectric material itself. These failures can degrade electrical performance, reduce the lifespan of microchips, and ultimately impact the overall functionality of electronic devices.

\begin{figure}[!t]
\centering
\includegraphics[width=3.0in]{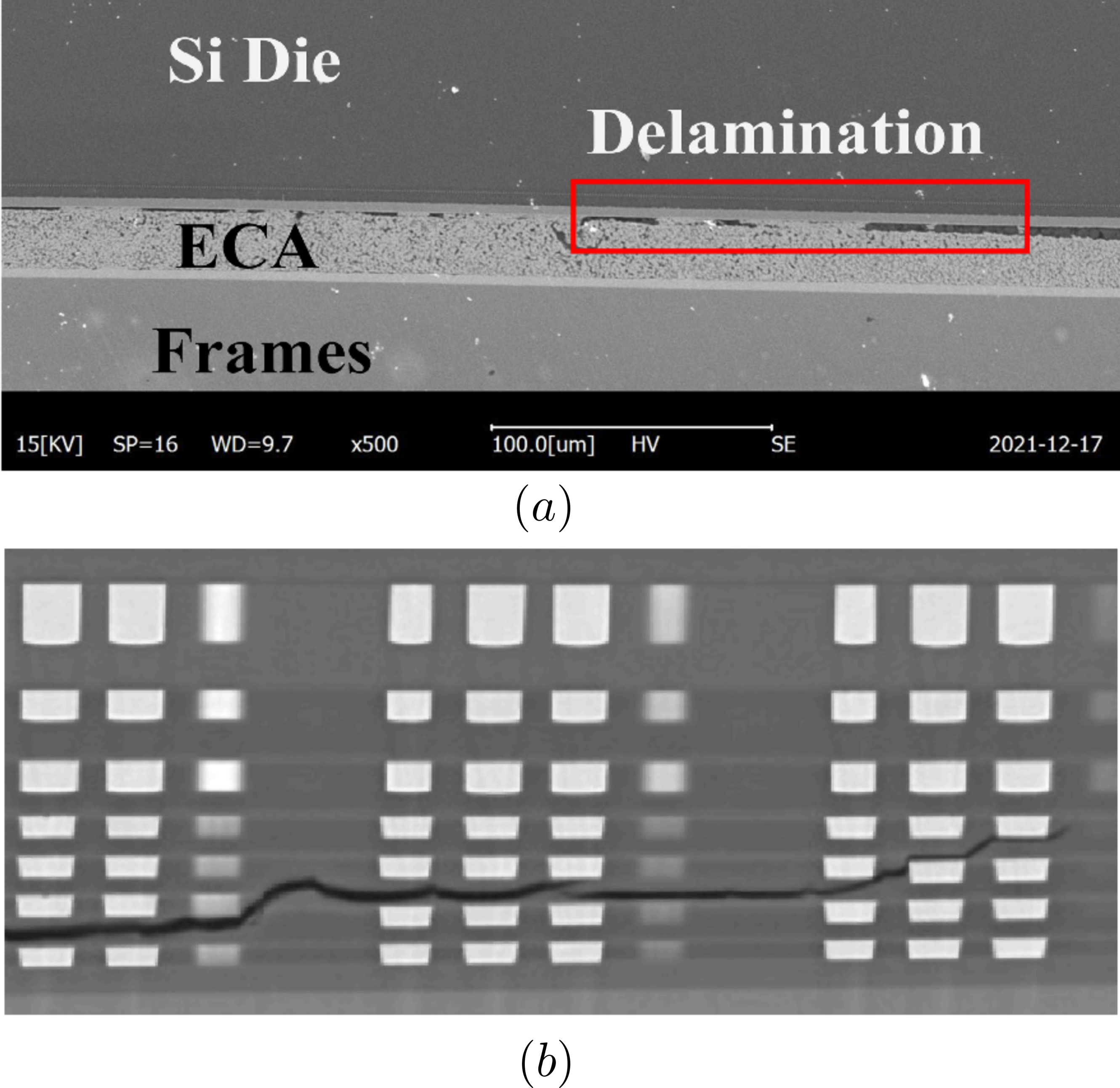}
    \caption{Illustration of delamination (adhesive failure) and fracture (cohesive failure):  (a). Illustration of delamination in a DFN-type Die/ECA-layered SEM diagram\cite{tian2024delamination}; (b). Illustration of fracture pattern in a Cu/ultra low-k multilevel interconnect structure\cite{zhang2009chip}. }
    \label{fig14: interconnect-failure-concept}
\end{figure}

Insulating ultra-low-k (ULK) materials, characterized by their low Young’s modulus and fracture toughness, compromises the mechanical integrity of the back-end-of-line (BEoL) stack. This structural weakness, combined with thermomechanical stresses induced by Chip Package Interaction (CPI), heightens the risk of failure. Specifically, delamination along Cu/dielectric interfaces (adhesive failure, see Figure.\ref{fig14: interconnect-failure-concept}(a)) and dielectric fracture (cohesive failure, see Figure.\ref{fig14: interconnect-failure-concept}(b)) are more likely to happen under these conditions. To mitigate fracture risks within the BEoL stack, both material selection and structural design must be optimized. One key approach involves the implementation of crack stop structures, which are specially designed, high-metal-density regions with no electrical functionality. These structures serve to dissipate energy in a manner that effectively slows or even halts crack propagation, thereby enhancing the mechanical resilience of the BEoL stack.

In this numerical benchmark example, a two-dimensional Cu/Ultra low-k multilevel interconnect layout model is applied with mechanically induced external loads and thermally induced external loads to demonstrate that the proposed CEM is capable of capturing fracture behavior during the Chip Package Interaction (CPI) process. 

\subsubsection{Mechanically-induced loads}
\begin{figure}[!t]
\centering
\includegraphics[width=3.0in]{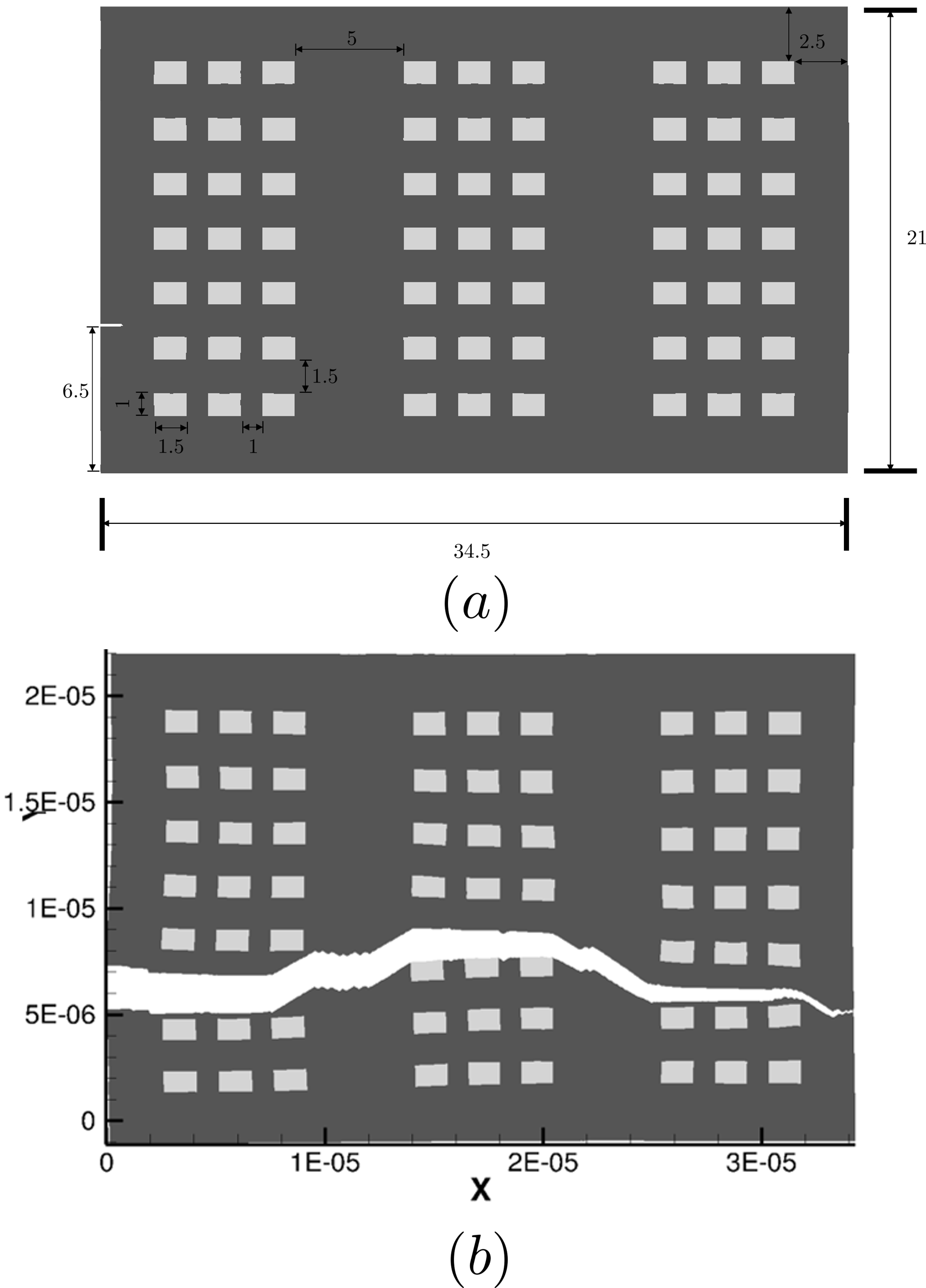}
    \caption{(a). The geometry of Cu/ultra low-k multilevel interconnect model (unit: $\mu m$); (b). Final crack propagation path. }
    \label{fig12: interconnect-mech-demo}
\end{figure}

The geometric dimensions of the Cu/Ultra low-k multilevel interconnect are shown in Figure.\ref{fig12: interconnect-mech-demo}(a). Two main materials are considered: copper (grey) and silicon dioxide (black). Due to a shortage of practical fab data, the layout of the multilevel two-dimensional structure is periodic, which is actually for the concept of proof rather than real Cu/Ultra low-k multilevel interconnect (see Figure.\ref{fig14: interconnect-failure-concept}(b)). 

The material properties are: copper density $\rho = 8960\ kg/m^3$, Young's modulus $E = 117\ GPa$, Possion ratio $\nu = 0.34$, critical fracture energy release rate $\mathcal{G}_c = 3.38\ J/m^2$; silicon dioxide density $\rho = 2270\ kg/m^3$, Young's modulus $E = 66.0\ GPa$, Poisson ratio $\nu = 0.17$, and critical fracture energy release rate $\mathcal{G}_c = 9.18\ J/m^2$. A small pre-notch exists at the left side of the multilevel structure to initiate a crack. The upper and bottom boundaries are prescribed with a velocity condition $v_0 = 200.0\ m/s$ to stretch the model. The total time of the process in simulation is $T = 1.0\times 10^{-8}s$. 

\begin{figure}[!t]
\centering
\includegraphics[width=3.5in]{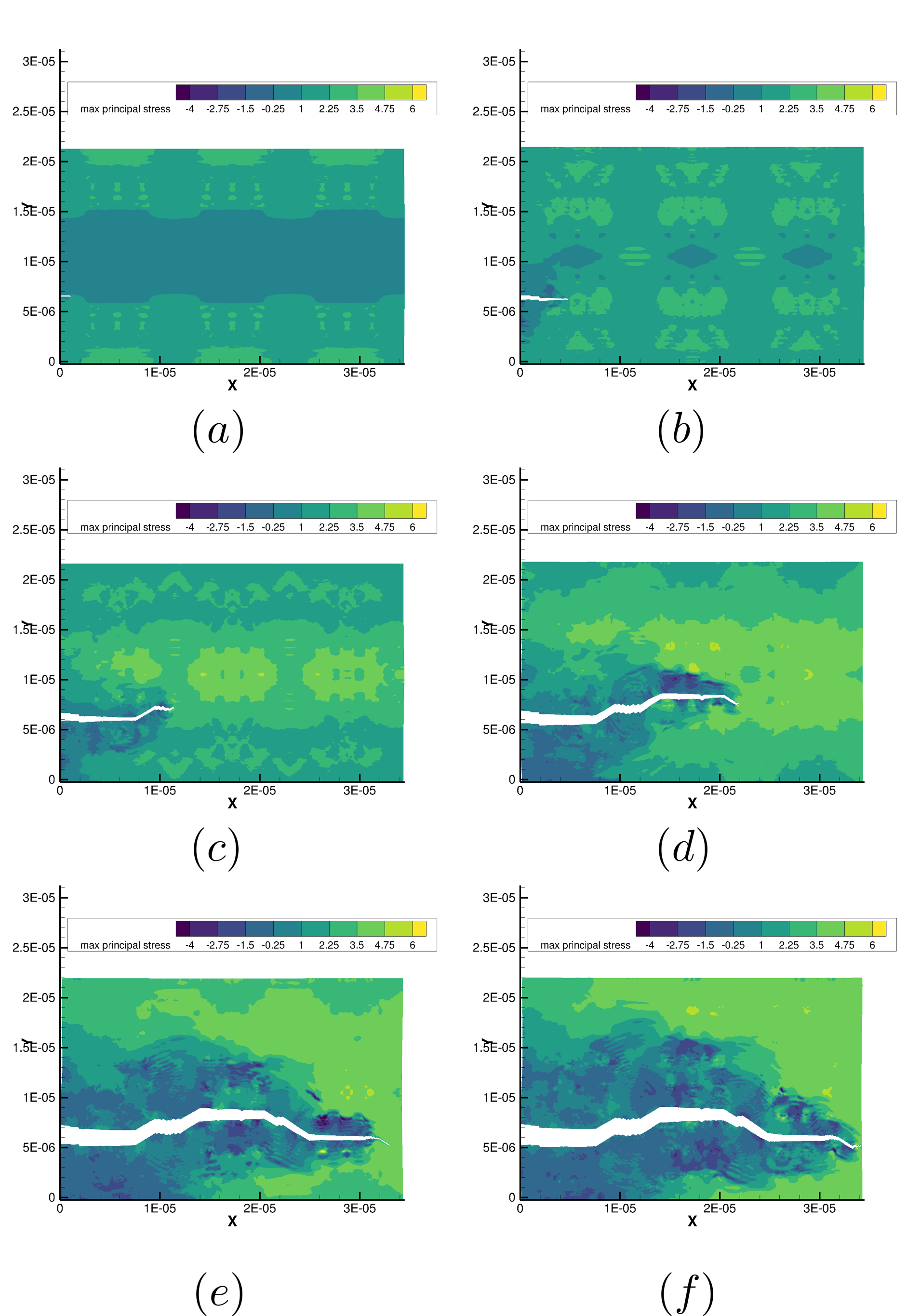}
    \caption{The maximum principal stress contour and crack propagation path evolution process: (a). time $t = 1.52\times 10^{-9} s$; (b). time $t = 2.519\times 10^{-9} s$; (c). time $t = 3.49\times 10^{-9} s$; (d). time $t = 4.517\times 10^{-9} s$; (e). time $t = 5.504\times 10^{-9} s$; (f). time $t = 5.858\times 10^{-9} s$. }
    \label{fig13: interconnect-mech-evolve}
\end{figure}

The final crack propagation pattern is shown in Figure.\ref{fig12: interconnect-mech-demo}(b). It is found that the crack pattern in the simulation model qualitatively agrees with the authenticating crack pattern in the fab (see Figure.\ref{fig14: interconnect-failure-concept}(b)). This simulation confirms the ability of the proposed CEM to model the fracturing process in a microelectronic device. Subsequently, the progressive crack propagation and its maximum principal stress fields are highlighted in Figure.\ref{fig13: interconnect-mech-evolve}(a-f).

\subsubsection{Thermally-induced loads}
The thermal effect on cracks in Cu/Ultra-low-k (ULK) multilevel interconnects is a critical reliability concern in advanced semiconductor technology. The primary issue arises from the mismatch in the coefficients of thermal expansion (CTE) between copper and ULK materials, which generates significant stress at their interface. During thermal cycling, such as wafer processing, packaging, and device operation, these stresses accumulate, leading to crack initiation and propagation.

\begin{figure}[!t]
\centering
\includegraphics[width=3.0in]{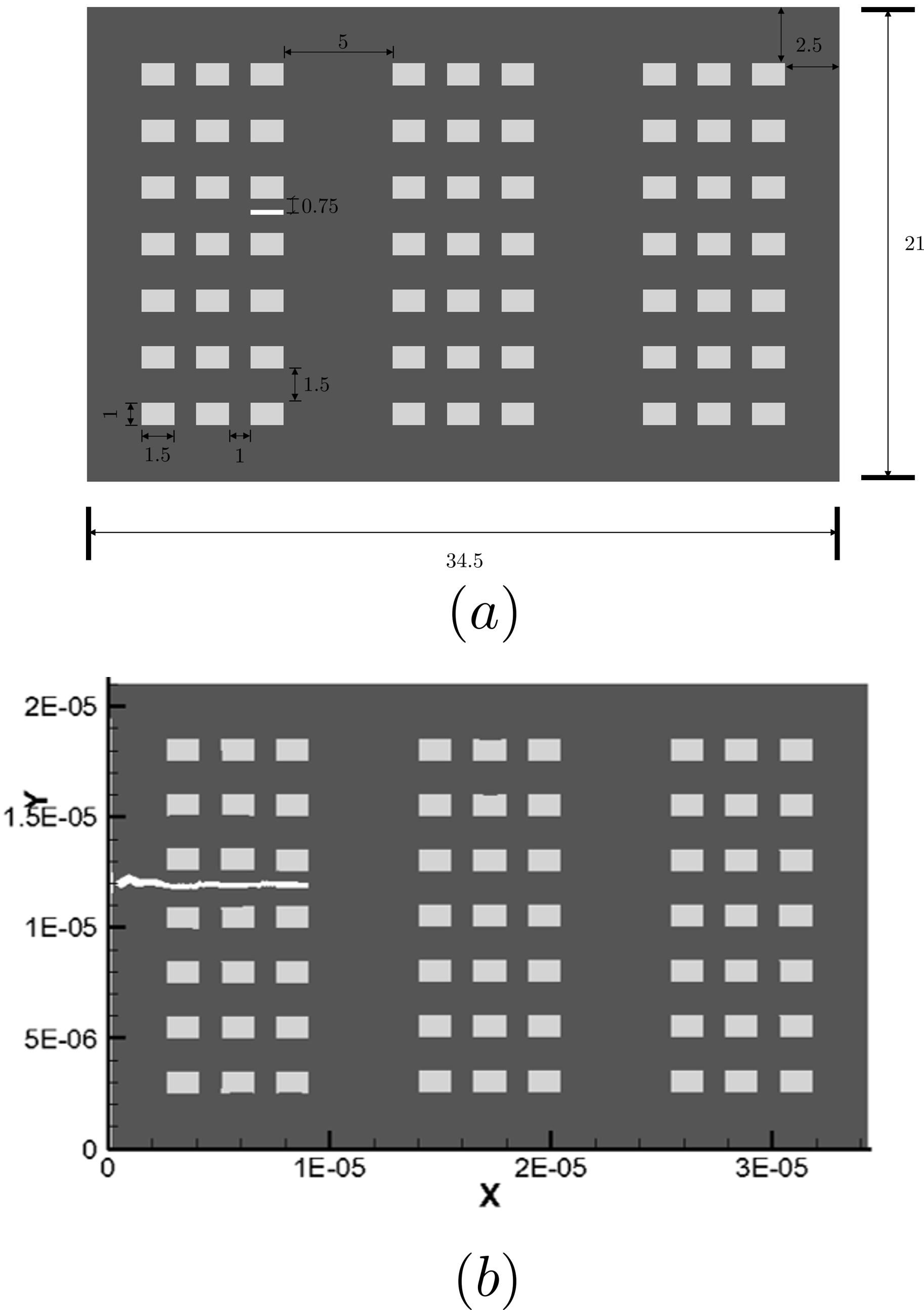}
    \caption{(a). The geometry of Cu/Ultra low-k multilevel interconnect model (unit: $\mu m$); (b). Final crack propagation path. }
    \label{fig15: interconnect-thermo-demo}
\end{figure}

Similarly, the geometric dimensions of the Cu/Ultra low-k multilevel interconnect model are shown in Figure.\ref{fig15: interconnect-thermo-demo}(a). Two materials are simulated: copper (grey) and silicon dioxide (black). The material properties including Young's modulus $E$, density $\rho$, Poisson ratio $v$, and critical fracture energy release rate $\mathcal{G}_c$ are the same as the mechanically-induced benchmark example. In addition to the material parameters above, thermal expansion coefficients for two materials are introduced here: copper of thermal expansion coefficient $\alpha = 16.5\times 10^{-6}\ K^{-1}$ and silicon dioxide of thermal expansion coefficient $\alpha = 0.5\times 10^{-6}\ K^{-1}$. A small pre-notch exists is inserted between two copper layers, and a crack is initiated here due to the substantial discrepancy of thermal expansion coefficients. The upper and lower boundaries are fixed. The temperature load is applied as follows:
\begin{equation}
dT =
\begin{cases} 
\frac{t}{t_0}\Delta T,  & \text{if }\ t \le t_0, \\
0, & \text{if } \ t > t_0.
\end{cases}
\end{equation}
in which, total temperature variation $\Delta T = -800\ K$, $t_0 = 8.0\times 10^{-10}s$. 

\begin{figure}[!t]
\centering
\includegraphics[width=3.5in]{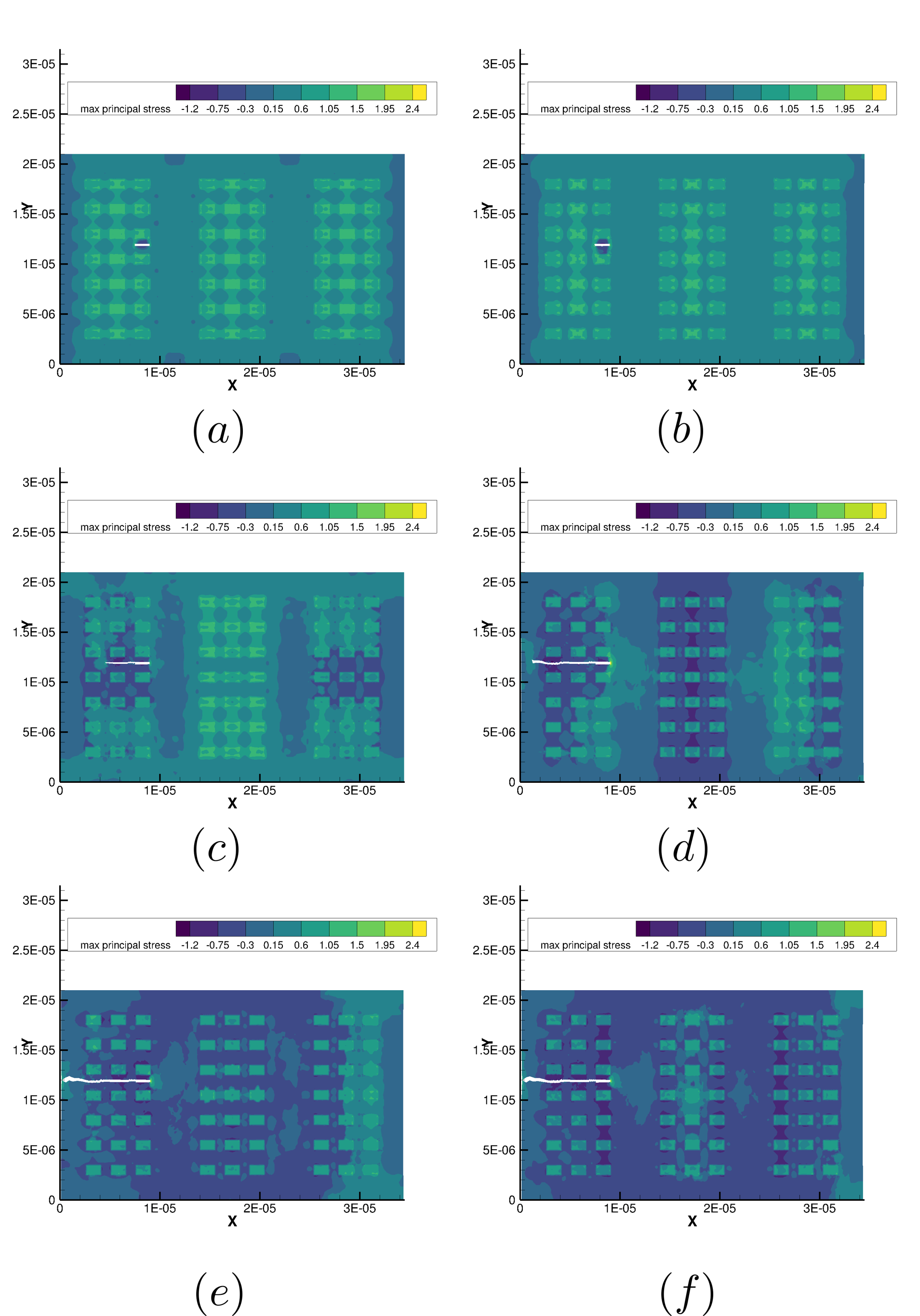}
    \caption{The maximum principal stress contour and crack propagation path evolution process: (a). time $t = 0.9\times 10^{-9} s$; (b). time $t = 1.158\times 10^{-9} s$; (c). time $t = 3.18\times 10^{-9} s$; (d). time $t = 5.4\times 10^{-9} s$; (e). time $t = 7.0\times 10^{-9} s$; (f). time $t = 8.0\times 10^{-9} s$. }
    \label{fig16: interconnect-thermo-evolve}
\end{figure}

The final crack propagation pattern is shown in Figure.\ref{fig15: interconnect-thermo-demo}(b). The crack initiates at left of the pre-notch and propagates parallel to the interconnect multilayer. Furthermore, the crack propagation progression and maximum principal stress fields are shown in snapshots in Figure.\ref{fig16: interconnect-thermo-evolve}(a-f). From the maximum principal stress contour, it is observed that the crack initiates at time $t = 1.158\times 10^{-9}s$ due to the obvious mismatch of coefficients of thermal expansion. It validates that the proposed CEM is capable of capturing thermally-induced crack propagation in Cu/Ultra low-k multilevel interconnect microelectronic devices.

\section{Conclusion}
A growing area of research is focused on mitigating Chip-Packaging Interaction, particularly in systems that incorporate stiff interconnects and brittle thin films. One promising strategy involves integrating metallic GR structures within the BEoL stacks. These GR structures absorb and dissipate externally induced stress energy, reducing the likelihood of crack initiation and propagation. Gaining a comprehensive understanding of the kinematics of dynamic fracture processes, especially in conjunction with preventive measures such as GR structures, is essential for enhancing the mechanical robustness and long-term reliability of on-chip interconnects. However, crack initiation and propagation under transient-dynamic loading conditions are inherently more complex than quasi-static loads. Dynamic effects introduce additional challenges such as inertia forces, stress wave interactions, crack branching, and fragmentation, all of which complicate the modeling and prediction of crack behavior. Moreover, stress singularities near crack tips, material heterogeneity, and variable boundary reflections further exacerbate this complexity, making dynamic fracture analysis critical and challenging in advanced microelectronic applications.

This study proposes a novel Crack Element Method developed within the Edge-based Smoothed Finite Element Method framework to address the inherent complexities of transient-dynamic fracture analysis. The CEM circumvents the limitations of conventional discrete and smeared crack modeling techniques by neither introducing additional nodes and element edges nor distributing damage across neighboring elements. This method minimizes sensitivity to mesh orientation and resolution, thereby enhancing computational efficiency, improving numerical stability, and allowing for a direct extension to three-dimensional domains without significant methodological reformulation. Central to the proposed method is a new failure criterion based on an element-level computation of the fracture energy release rate, intrinsically linked to the topology of the split elements. In contrast to traditional approaches, such as the J-integral, which suffer from mesh dependency, path integration ambiguity, and difficulty tracking rapidly evolving crack tips under dynamic loading, this criterion demonstrates improved objectivity and robustness. The energy-based formulation also mitigates numerical artifacts such as crack path trapping induced by stress wave interactions and high-frequency fluctuations.

The effectiveness of the proposed CEM is validated through a suite of numerical benchmarks. In the first three transient-dynamic fracture examples, the method accurately captures crack trajectories, dissipated fracture energy, and crack tip velocities, with yielding results that show strong agreement with experimental data and existing numerical solutions. Notably, in the two-offset-notch benchmark, the CEM exhibits exceptional capability in resolving complex crack tip interactions and double stress singularities, despite its formulation based solely on mode-I fracture mechanics, underscoring its physical fidelity and versatility.

Finally, a semiconductor-related case study involving Cu/ultra-low-k interconnect structures is conducted to evaluate the method’s practical use in microelectronic reliability analysis. The accurate prediction of failure mechanisms caused by mechanical and thermal factors confirms the method’s potential as a robust simulation tool for advanced packaging technologies. The proposed finite element modeling method offers a dynamic, locally adaptive algorithm capable of efficiently and accurately modeling crack evolution in quasi-brittle materials under transient loading, making it as a strong alternative to existing computational fracture mechanics frameworks. Nevertheless, a comprehensive three-dimensional extension of the method remains to be addressed. This includes the simulating crack branching phenomena and implementing GPU-accelerated computation for practical industrial use, which will be presented in a forthcoming paper.




 
\bibliography{IEEE-refs}
%

\bibliographystyle{IEEEtran}











\newpage

 



\begin{IEEEbiographynophoto}{Yuxi Xie}
received the B.S. degree in civil engineering from Huazhong University of Science and Technology, Wuhan, China, in 2014, and the Ph.D. degree in civil engineering from University of California at Berkeley, Berkeley, CA, USA, in 2021. He is currently a R\&D engineer in Synopsys Inc. His research interests include computational fracture mechanics, microscopic mechanics, additive manufacturing, and scientific machine learning. 
\end{IEEEbiographynophoto}

\begin{IEEEbiographynophoto}{Ethan J. Wu}
is a senior at Granada High School. He has been involved in this research project since the summer of his sophomore year, focusing on the element-splitting algorithm and the design of topological element structures to track crack growth. His research interests include geometric mathematics, biochemistry, molecular biology, and machine learning, with applications in bio-chip technologies for biomedical use.
\end{IEEEbiographynophoto}

\begin{IEEEbiographynophoto}{Lu Xu}
is the Director for 18\AA business line at Intel, where he leads profit \& loss strategy, business planning, and go-to-market execution. Dr.Xu holds a B.Sc. in Materials Science from Shanghai Jiao Tong University, China, and a Ph.D. in Electronic Engineering from Dublin City University, Ireland. His research interests include lithography, multi-physics simulation, device physics, and design flows for semiconductor foundries.
\end{IEEEbiographynophoto}

\begin{IEEEbiographynophoto}{Jimmy Perez}
received the B.S. degree in mechanical engineering from the University of Puerto Rico - Mayag\"uez, and an M.S. in Manufacturing Systems Engineering from the University of Wisconsin-Madison in 1998. He is currently an Area Manager at Intel Corporation on controlled collapse chip connection (C4) with a focus on metrology, defect detedction, and NPI.
\end{IEEEbiographynophoto}

\begin{IEEEbiographynophoto}{Shaofan Li}
received the B.S. degree in Mechanical Engineering from the East China University of Science and Technology in 1982 and he received the Ph.D. degree in Mechanical Engineering from Northwestern University in 1997. Dr. Shaofan Li is currently a full professor at the University of California at Berkeley. He has been working on engineering mechanics and machine learning neural networks.
\end{IEEEbiographynophoto}

\vfill

\end{document}